\documentclass[lettersize,journal]{IEEEtran}
\usepackage{amsmath,amsfonts}
\usepackage{algorithmic}
\usepackage{algorithm}
\usepackage{array}
\usepackage[caption=false,font=normalsize,labelfont=sf,textfont=sf]{subfig}
\usepackage{textcomp}
\usepackage{stfloats}
\usepackage{url}
\usepackage{verbatim}
\usepackage{graphicx}
\usepackage{cite}
\usepackage{amsmath,graphicx,epsfig,multirow,adjustbox,booktabs,caption,color,amssymb}
\usepackage{lipsum}
\usepackage{fancyhdr}
\usepackage{mathtools}

\usepackage{color}

\newcommand{\tabincell}[2]{\begin{tabular}{@{}#1@{}}#2\end{tabular}}

\begin{document}

\title{TMS: A Temporal Multi-scale Backbone Design for Speaker Embedding}

\author{Ruiteng Zhang$^{1}$, Jianguo Wei$^{1,2}$, Xugang Lu$^{3}$, Wenhuan Lu$^{1}$, Di Jin$^{1}$, Junhai Xu$^{1}$, Lin Zhang$^{4}$, Yantao Ji$^{5}$, Jianwu Dang$^{1}$
\thanks{1. College of Intelligence and Computing, Tianjin University, Tianjin, China.}
\thanks{2. Computer College, Qinghai Nationalities University, Xining, China.}
\thanks{3. National Institute of Information and Communications Technology, Kyoto, Japan.}
\thanks{4. National Institute of Informatics, Tokyo, Japan.}
\thanks{5. School of Software Engineering, Xi’an Jiaotong University, Xi’an, China.}
\thanks{Corresponding author: Wenhuan Lu.}
}



\maketitle

\begin{abstract}
Speaker embedding is an important front-end module to explore discriminative speaker features (e.g., X-vector) for many speech applications where speaker information is needed. Current state-of-the-art backbone networks for speaker embedding are designed to aggregate multi-scale features from an utterance with multi-branch network architectures for speaker representation (e.g., ECAPA-TDNN). However, naively adding many branches of multi-scale features with the simple fully convolutional operation could not efficiently improve the performance due to the rapid increase of model parameters and computational complexity. Therefore, in the most current state-of-the-art network architectures, only a few branches corresponding to a limited number of temporal scales could be designed for speaker embeddings. To address this problem, in this paper, we propose an effective temporal multi-scale (TMS) model where multi-scale branches could be efficiently designed in a speaker embedding network almost without increasing computational costs. The new model is based on the conventional time-delay neural network (TDNN), where the network architecture is smartly separated into two modeling operators: a channel-modeling operator and a temporal multi-branch modeling operator. Adding temporal multi-scale in the temporal multi-branch operator needs only a little bit increase of the number of parameters, and thus save more computational budget for adding more branches with large temporal scales. Moreover, after the model was trained, in the inference stage, we further developed a systemic re-parameterization method to convert the multi-branch network topology into a single-path-based topology in order to increase inference speed. We investigated the performance of the new TMS method for automatic speaker verification (ASV) on in-domain (VoxCeleb) and out-of-domain (CNCeleb) conditions. Results show that the model based on the TMS method obtained a significant increase in the performance over the state-of-the-art ASV models, i.e.,  ECAPA-TDNN, and meanwhile, had a better model generalization. Moreover, the proposed model achieved a 29\% -- 46\% speed up in inference compared to the state-of-the-art ECAPA-TDNN.
\end{abstract}

\begin{IEEEkeywords}
Speaker verification, Temporal multi-scale feature, Effective backbone.
\end{IEEEkeywords}

\section{Introduction}
\IEEEPARstart{S}{peaker} embedding is widely used as a front-end processing for speaker discriminative information extraction for speech application systems where speaker information is needed, for example, speaker verification systems in authentication for security access \cite{XU2020394}, speaker diarization systems in real-time meeting recordings and/or dialogs \cite{tranter2006overview, horiguchi2020end}. Due to the success of deep learning frameworks in speech and image processing, speaker embedding algorithms have been proposed in which outputs of bottleneck layers could be used as speaker representation. The early speaker embedding algorithms were proposed based on deep neural network (DNN), where d-vectors were extracted from a bottleneck layer of the DNN for speaker representation \cite{d-vector}. Later, some most successful speaker embedding algorithm was proposed based on time-delay neural network (TDNN) \cite{9376629}, where X-vectors \cite{x-vector} were extracted through statistical pooling to convert utterances with various durations into fixed dimension speaker vectors. Inspired by the success of X-vector, several new speaker embedding algorithms have been proposed with manipulation of network architectures, for example, extended-TDNN (E-TDNN) \cite{e-tdnn}, factorized-TDNN (F-TDNN) \cite{povey2018semi}, and ResNet \cite{resnet}. 
Also, several studies tried to modify the learning objective functions to improve the speaker embedding algorithms, for example, A-Softmax \cite{a-softmax}, AM-Softmax \cite{am-softmax}, AAM-Softmax \cite{arc-softmax}, etc.
Recently, by combining the advantages of the model architecture and learning objective functions, ECAPA-TDNN as a new speaker embedding model (with Res2Net module \cite{gao2019res2net} and attention \cite{hu2018squeeze}) has achieved the state-of-the-art performance on VoxCeleb test set for automatic speaker verification (ASV) task \cite{zhou2021resnext, ecapa-tdnn}. The main finding of all above-mentioned studies confirm that integrating multi-scale features of speech could consistently improve the performance for speaker feature extraction for ASV. This finding is consistent with the physiological study that speaker characteristics are encoded in multiple temporal scales of acoustic speech with short- and long-term segments (e.g., phonetics, prosody, etc.) \cite{laver1994principles}.   
In order to explore these multi-scale features, either single-path or multi-branch network could be designed.

\subsection{Single-path-based speaker embedding models}

As the most famous speaker feature, X-vector representation has been successfully applied in ASV \cite{x-vector}. This X-vector was extracted from a well-trained speaker embedding model based on TDNN \cite{9376629}. In the speaker embedding model, TDNN extracted the frame-level features through the stacking of several TDNN layers where the temporal contextual relationship of adjacent frames was taken into consideration. Moreover, a pooling operator aggregates frame-level features to segment-level features, which were passed to bottleneck layers (fully connected (FC) layers) for further feature extraction. Speaker embeddings were extracted from these bottleneck FC layers. This classical design lays the foundation for most of the follow-up ASV systems that could keep robust in various real-world conditions. 
Because of the high performance of X-vector systems, further studies tried to develop improved TDNN models to enhance the deep representations for the ASV system, for example, E-TDNN \cite{e-tdnn},  F-TDNN \cite{povey2018semi}, D-TDNN \cite{yu2020densely}, and ARET \cite{zhang2020aret} etc.
Although the model architectures were designed as single-path topology, the multiple scales of features could be explored through the stacking of several layers (deep structure), therefore, multiple scales of features could be extracted from acoustic signals for speaker representations. 

\subsection{Multi-branch-based speaker embedding models }

Although the single-path-based speaker embedding models could achieve superior performance, they are lack capability and efficiency in passing multi-scale features for final classifier modeling. In computer vision (CV), recognizing one object needs to consider the object itself and model the different sizes of contextual information simultaneously \cite{gao2019res2net}. 
In speech, multi-scale features also exist in acoustic signals for recognition tasks. In one aspect, speech has multi-scale and hierarchical linguistic structures, e.g., phoneme, syllable, word, etc. In another aspect,  different speech production organs have different time-frequency responses that contribute to acoustic features during dynamic movements \cite{kitamura2005individual, takemoto2006acoustic}. These two aspects determine that the speaker features should be encoded in short- and long-term segments of speech signals that reflect the physiological property of speech production organs, speaking styles, changes of prosody, and so on. Therefore, local and global representations should be emphasized for  speaker discriminative information exploration, i.e., modeling for temporal multi-scale feature extraction.

\begin{figure*}[ht]
\setlength{\abovecaptionskip}{2pt}
\begin{minipage}[t]{0.08\linewidth}
  \centering
  \centerline{\epsfig{figure=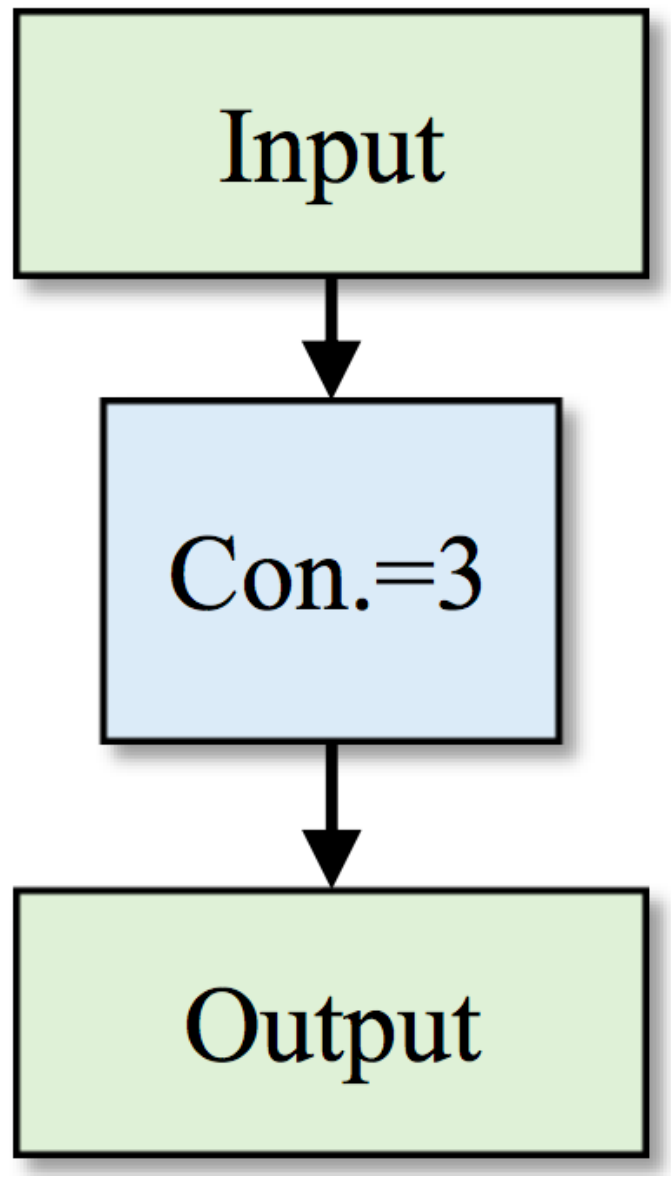,width=1.3cm}}
  \centerline{{(a)}}\medskip
\end{minipage}
\begin{minipage}[t]{0.2\linewidth}
  \centering
  \centerline{\epsfig{figure=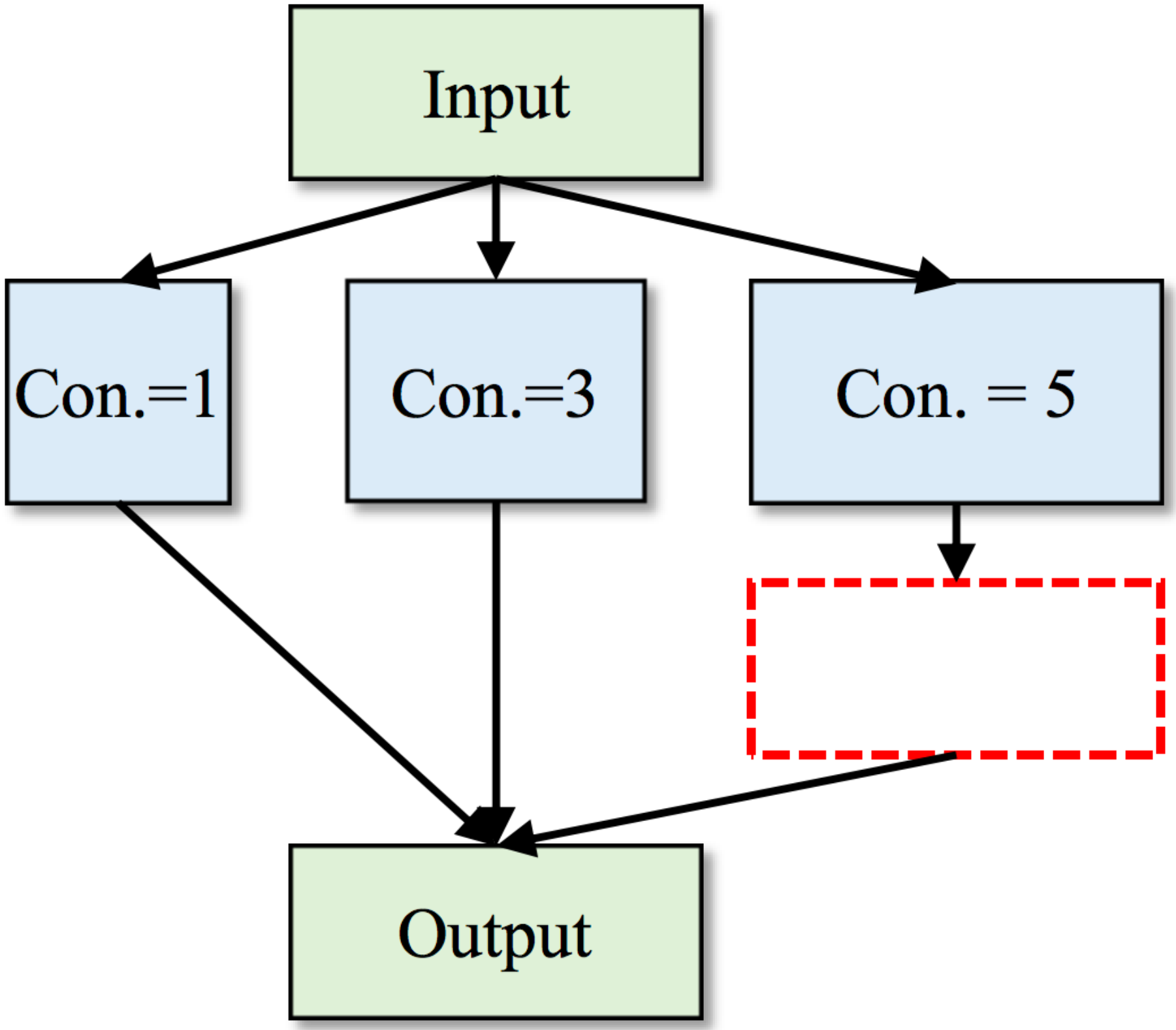,width=3.5cm}}
  \centerline{{(b)}}\medskip
\end{minipage}
\begin{minipage}[t]{0.19\linewidth}
  \centering
  \centerline{\epsfig{figure=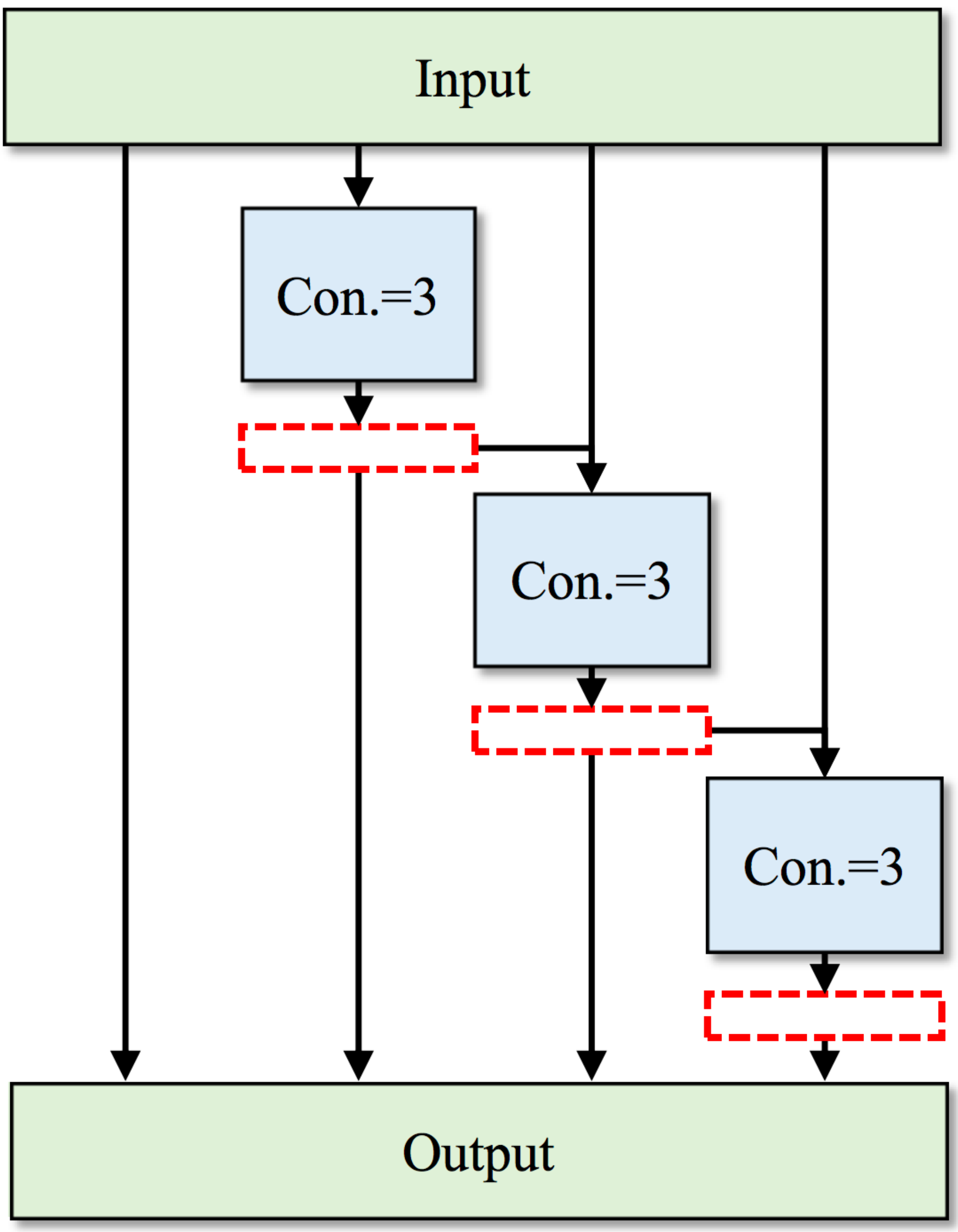,width=3.3cm}}
  \centerline{{(c)}}\medskip
\end{minipage}
\begin{minipage}[t]{0.2\linewidth}
  \raggedleft
  \centerline{\epsfig{figure=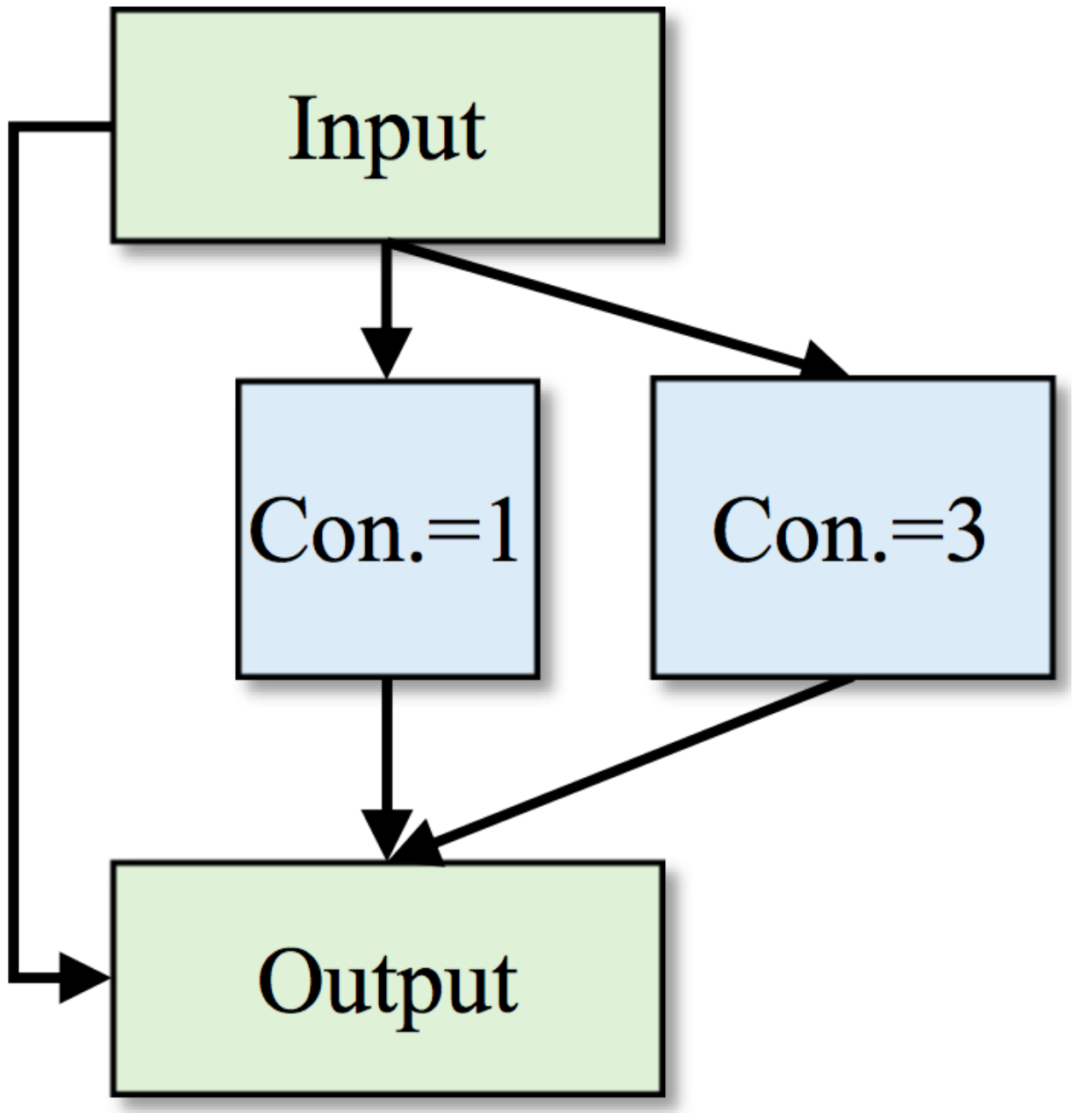,width=2.3cm}}
  \centerline{{(d)}}\medskip
\end{minipage}
\begin{minipage}[t]{0.27\linewidth}
  \centering
  \centerline{\epsfig{figure=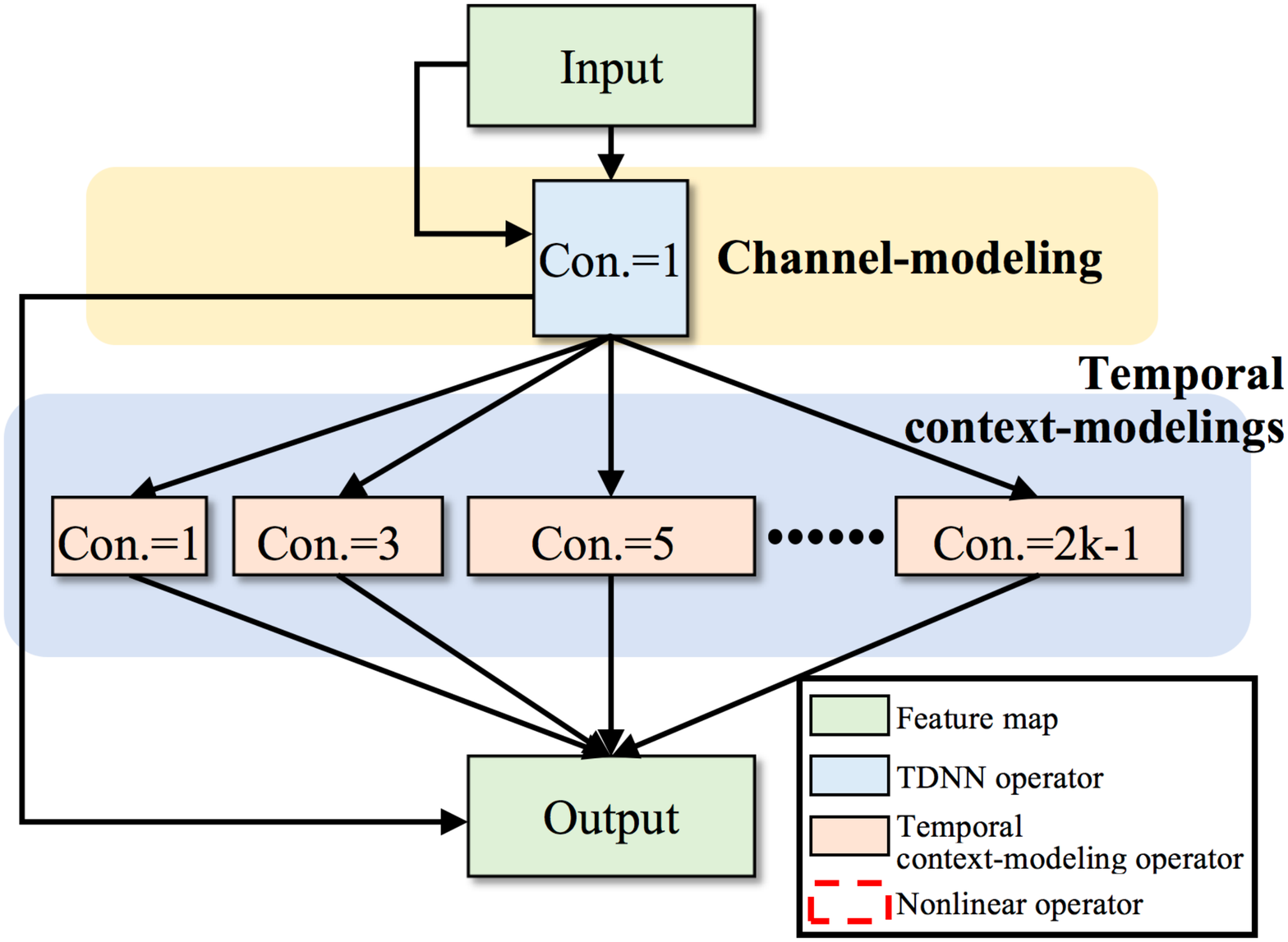,width=6cm}}
  \centerline{{(e)}}\medskip
\end{minipage}
\caption{Topology comparisons in speaker embedding networks: (a) TDNN, (b) InceptionNet, (c) Res2Net, (d) RepVGG, (e) the proposed {\bf temporal multi-scale (TMS)} module. Con. is the size of temporal context (the kernel size of the 1-D convolutional operation), and $k$ is the branch index (as used in (e)).}
\label{fig:idea_diagram}
\end{figure*}

The original multi-scale feature models were proposed in the CV field using a multi-branch design with different convolutional kernels, for example, InceptionNets  \cite{szegedy2017inception}, DenseNet \cite{huang2017densely}, and ResNest \cite{zhang2020resnest}. In speech, there is also a model architecture design paradigm shift from single-branch to multi-branch models in order to capture multi-scale speech features. This model architecture paradigm shift with network topology comparisons is illustrated in Fig. 1. In this figure, the TDNN module only has a single-path topology with 1-D convolutional that kernel size = 3 (Fig. \ref{fig:idea_diagram} (a)). In Fig. \ref{fig:idea_diagram} (b), as proposed in InceptionNet \cite{szegedy2017inception}, multi-parallel branches with different convolutional kernels and nonlinear operators are designed to explore multi-scale features (scale size 1, 3, and 5 as in this example with three different convolutional kernels). The Res2Net-based architecture is integrated into the multi-branch design \cite{gao2019res2net}, as shown in Fig. \ref{fig:idea_diagram} (c). This approach is based on the idea of ``making a residual connection in the residual connection” \cite{gao2019res2net}. Representative studies of this designs are Res2Net \cite{gao2019res2net}, HS-Net \cite{yuan2020hs}, and ECAPA-TDNN \cite{ecapa-tdnn}. In particular, ECAPA-TDNN \cite{ecapa-tdnn} is the state-of-the-art speaker embedding model used in ASV with a multi-branch topology reference to Res2Net \cite{gao2019res2net}, and has achieved an excellent performance in recent ASV competitions and the industry \cite{voxsrc_ecapa_1, voxsrc_ecapa_2, zeinali2020sdsv, zeinali2019short}. 
However, as studies showed that this multi-branch (e.g., Res2Net structure) is not friendly for parallel computing \cite{ding2021repvgg}. The main reason is that the multi-branch designs increase the memory access cost (MAC) \cite{ma2018shufflenet}. This disadvantage makes it difficult  to deploy the Res2Net-based model for real applications.

In order to increase the parallelization ability of models, re-parameterized parallel multi-branch structure has been designed (Fig. \ref{fig:idea_diagram} (d)) \cite{ding2021repvgg}. It supports network backbones with different topologies between training and inference stages. In the training stage, the model adopts the multi-branch topology to learn the multi-scale speaker's features, while the network uses the re-parameterization approach to convert all branches to a single-path topology in calculation with a high inference speed in the inference stage. The idea of re-parameterization was first proposed in \cite{ding2021repvgg} on RepVGG model for image processing, and was later adapted to ASV in \cite{zhang2021cs, ma2021rep}. But these models were based on Rep-VGG's structure, where only a limited number of branches (temporal scales) could be integrated. 
Naively adding many branches in those model architectures will add excessive training parameters and make models hard to converge with increasing large computational load in model training. In addition, in the re-parameterization strategy, the combination of branches must adapt to the largest kernel size of the branch, which makes the computational load strongly rely on the branch with the largest kernel size.

\subsection{Our focus and contributions}

Considering the problems mentioned above, in this paper, we propose an effective speaker embedding model with integrating temporal multi-scale (TMS) processing in multi-branch neural network architecture and extracting speaker features for ASV. 
The basic TMS module of the proposed model is illustrated in Fig. \ref{fig:idea_diagram} (e). 
In this model, the TMS has much more temporal multi-scale branches each of which has with much wider temporal receptive fields than other models in previous studies.
With this specially designed module, there is no large increase in computational cost, although there is a significant improvement to capture multi-scale speaker features.
Moreover, based on the re-parameterization method, the TMS could be easily converted to a single-path topology hence increasing the inference speed. 
This proposed TMS can be used as an out-of-the-box module, which could be conveniently adopted for all types of TDNN layers (TMS-TDNN). 
Based on the TMS-TDNN, we designed a re-parameterized attention-based TMS-TDNN (Rep-A-TMS-TDNN) model for final ASV experiments. 
Experimental results show that the proposed TMS module is able to decrease the computational cost of the conventional multi-branch TDNN layer while increasing verification accuracy. In particular, the performance of the proposed Rep-A-TMS-TDNN can surpass the state-of-the-art ASV models (e.g., ECAPA-TDNN) in both the verification accuracy and inference speed. In summary, our contributions are as follows: 
\begin{enumerate}
\item We proposed a model design strategy called TMS to effectively integrate multi-scale speaker feature extraction in speaker embedding model. Different from most studies (that integrate multi-scale feature extraction by using different kernel sizes with the fully convolutional operation), our strategy only considers the multi-scale in the temporal dimension with independent processing of channel and temporal processing. The advantage of our proposal is that we could easily incorporate a large number of branches with a large size of temporal scales in speaker embedding model with an ignorable increase in computational cost. 
\item We developed a systemic re-parameterization method to convert the TMS module to a single-path topology, which can increase the inference speed, and meanwhile, bringing almost negligible influence in verification accuracy. 
\item We made a detailed and comprehensive analysis and experiments with the ASV task on the in-domain (VoxCeleb) and out-of-domain (CNCeleb) conditions, and confirmed the effectiveness of the proposed TMS in speaker embedding modeling. 
\end{enumerate}

The remainder of the paper is organized as follows. Section II introduces the proposed TMS strategy and the re-parameterization strategy, as well as  our model implementations. Section III examined the model with ASV experiments. Section IV further checks several factors that may affect the performance in ASV. Conclusions and future works are given in Section V.

\section{Proposed Temporal Multi-scale Strategy}

\begin{figure*}[ht]
\setlength{\abovecaptionskip}{2pt}
\begin{minipage}[t]{0.3\linewidth}
  \centering
  \centerline{\epsfig{figure=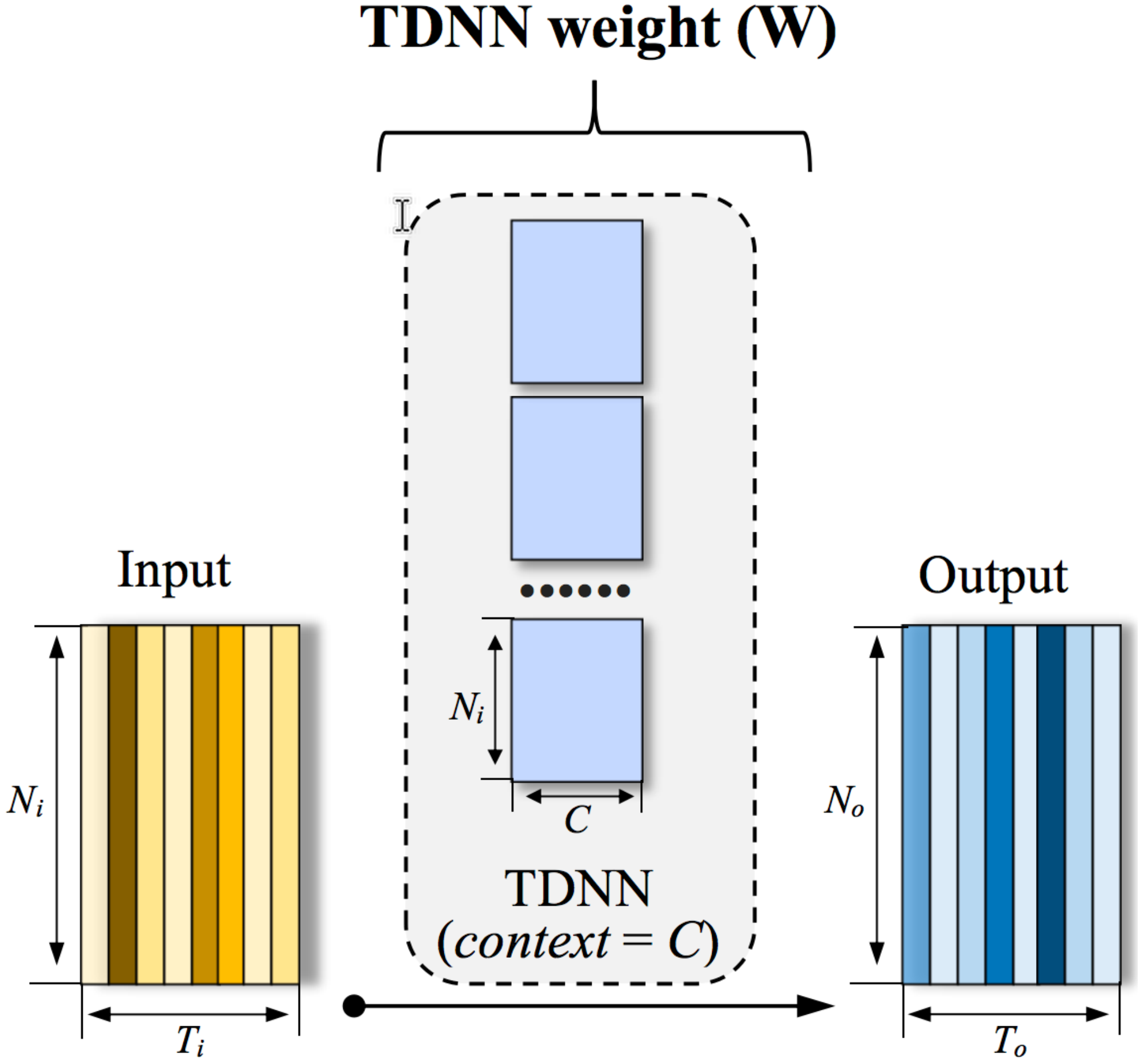,width=5.3cm}}
  \centerline{{(a)}}\medskip
\end{minipage}
\begin{minipage}[t]{0.7\linewidth}
  \centering
  \centerline{\epsfig{figure=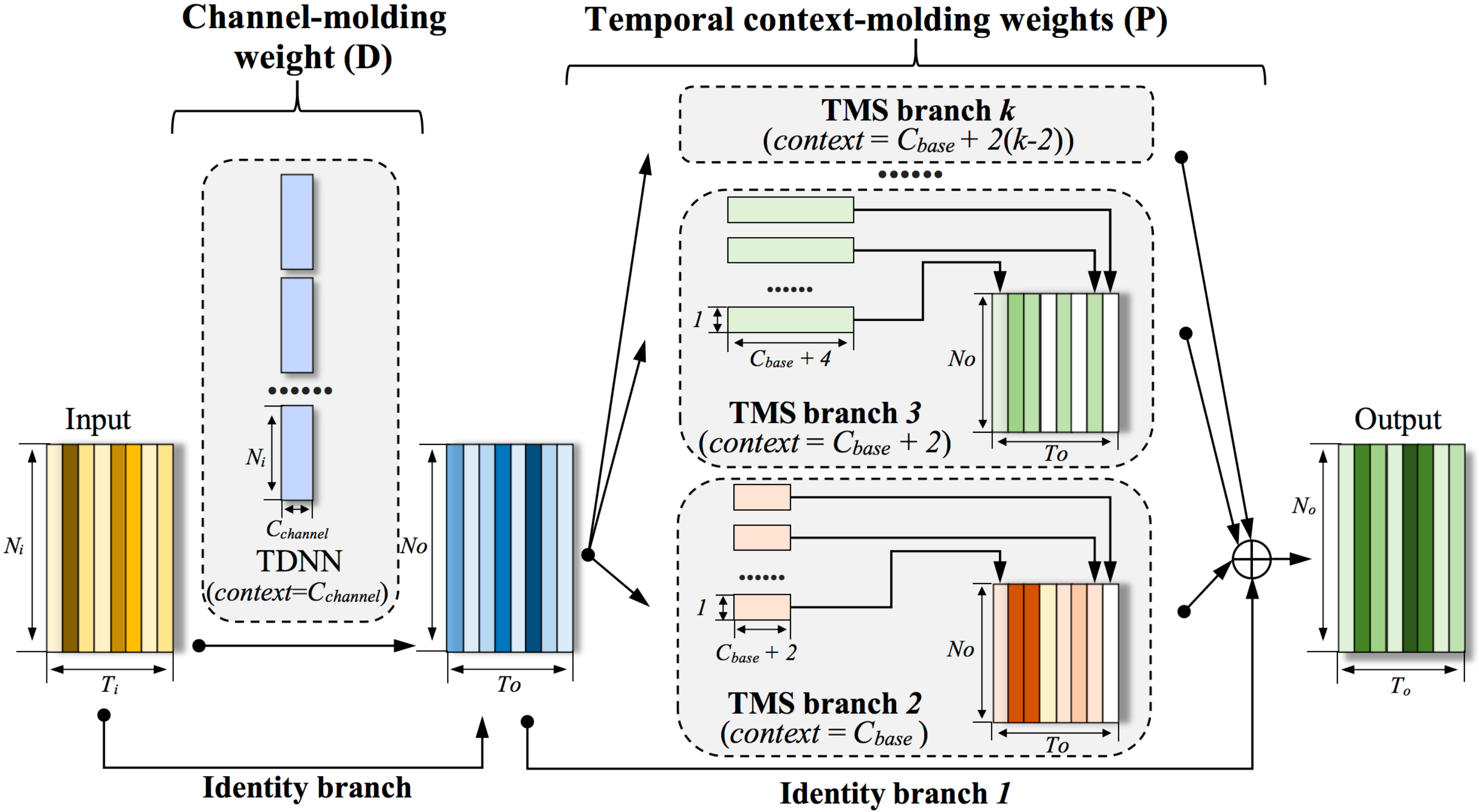,width=12cm}}
  \centerline{{(b)}}\medskip
\end{minipage}
\begin{minipage}[t]{0.9\linewidth}
  \centering
  \centerline{\epsfig{figure=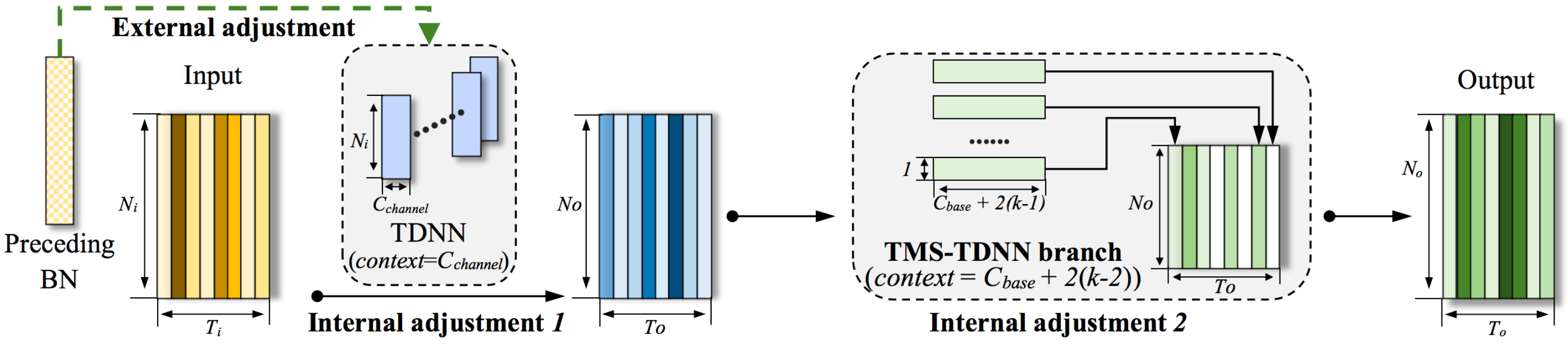,width=14cm}}
 \centerline{{(c)}}\medskip
\end{minipage}
\caption{Network diagram for (a) TDNN, (b) TMS, and (c) the re-parameterized TMS. The ``context" in this figure denotes the temporal context of TDNN.}
\label{fig:cdm_diagram}
\end{figure*}

The general sketch of the proposed TMS module was introduced in Fig. \ref{fig:idea_diagram} (d), and the detailed TMS network architecture design and re-parameterization process were illustrated in Fig. \ref{fig:cdm_diagram}. In this section, we will introduce them in detail.

\subsection{The proposed TMS network architecture}

The proposed TMS architecture is shown in Fig. \ref{fig:cdm_diagram} (b). For the convenience of explanation, the conventional TDNN based operation is also shown in Fig. \ref{fig:cdm_diagram} (a) for comparison. As shown in Fig. \ref{fig:cdm_diagram} (b), in the TMS, we first decouple the traditional TDNN operator into two parts: the channel-modeling operator and TMS-modeling operator. The channel-modeling operator focuses on modeling the channel relationship for the features, which can be developed by a common TDNN operator, but their temporal context could be rigorously controlled and is typically much more narrower than that used in a regular TDNN operator (Fig. \ref{fig:cdm_diagram} (a) and Fig. \ref{fig:cdm_diagram} (b) shows the comparison). Then, the TMS-modeling part includes a series of branches that independently consider temporal contextual relationships of every input channel, helping the network to explore local and global speaker features sufficiently with little increase of computational costs. Finally, all outputs of the TMS branches are added to be one tensor that serves as the output of the TMS. Compared to traditional multi-branch networks, the TMS strategy could support more branches and include longer contextual information of one branch.

A typical TDNN's weight is represented as ${\bf W} \in \mathbb{R}^{N_{o} \times N_{i} \times C}$, where $N_{i}$ and $N_{o}$ are the number of input and output channels of the TDNN, and $C$ is the temporal context of the weight of each channel. 
${\bf X} \in \mathbb{R}^{N_{i} \times T_{i}}$ is the input tensor and ${\bf Y} \in \mathbb{R}^{N_{o} \times T_{o}}$ is the output tensor, where $T_{i}$ and $T_{o}$ are the numbers of  frames of input and output.
For more details, in the typical TDNN (Fig. \ref{fig:cdm_diagram} (a)), the output ${\bf Y}$ is calculated as:
\begin{eqnarray}
\begin{aligned}
{\bf Y} &= {\bf W} * {\bf X} \\
{\bf y}_{n_{o},:} &= \sum\limits_{n_{i}=1}^{N_{i}} {\bf W}_{n_{o},:,:} * {\bf x}_{n_{i},:} \quad,
\end{aligned}
\end{eqnarray}
where ``*'' is the discrete convolution operator, ${\bf y}_{n_{o},:}$ is the tensor slice along the ${n_{o}}$-th output channel of ${\bf Y}$, $n_o \in \{1,2,...,N_o\}$. ${\bf W}_{n_{o},:,:}$ is the tensor slice of ${\bf W}$, and ${\bf x}_{n_{i},:}$ is the tensor slice of ${\bf X}$.

Conventional multi-scale networks often adopted the fully convolutional operator (weight is consisted of a ${N_{o} \times N_{i} \times C}$ tensor) to design a parallel multi-branch topology. 
The heavy computational load of this structure limited the usage of kernels with a large size and number of branches. For example, in RepVGG, it is hard to use a branch with a large convolutional kernel with the size of 7$\times$7. 
Therefore, it is difficult for RepVGG to explore long-segmental speaker features. Based on these considerations, two disadvantages limit the performance of the conventional multi-branch models, i.e., 
(1) the fewer number of branches limited the modeling capability, and (2) the sample and small receptive fields of the convolutional kernels can not efficiently explore those complex multi-scale speaker information.

Unlike conventional parallel multi-branch models, the TMS strategy first separates TDNN to a channel-modeling operator and a temporal context-modeling operator (Fig. \ref{fig:cdm_diagram} (b)). The channel-modeling operator with a rigorously controlled temporal context is responsible for modeling the channel information. And the context-modeling operator can model the temporal contextual information with a little bit of the increase of the computational cost. The channel-modeling operator is:
\begin{eqnarray}
\label{eq:channel-modeling}
\begin{aligned}
{\bf Y} &= {\bf D} * {\bf X} \\ 
{\bf y}_{n_{o},:} &= \sum\limits_{n_{i}=1}^{N_{i}}  {\bf D}_{n_{o},:,:} * {\bf x}_{n_{i},:} \quad,
\end{aligned}
\end{eqnarray}
where ${\bf D} \in \mathbb{R}^{N_{o} \times N_{i} \times C_{channel}}$, $C_{channel}$ is the temporal context of the channel-modeling operator, and ${\bf D}_{n_{o},:,:}$ is the tensor slice with the ${n_{o}}$-th output channel of the channel-modeling's weight ${\bf D}$.
The temporal context-modeling operator is:
\begin{eqnarray}
\begin{aligned}
\label{eq:context-modeling}
{\bf Y'} &= {\bf P} * {\bf X'}\\
{\bf y'}_{n_{o},:} &= {\bf P}_{n_{o},:,:} * {\bf x'}_{n_{i},: }\quad, 
\end{aligned}
\end{eqnarray}
where ${\bf P} \in \mathbb{R}^{N_{o} \times N_{i} \times C_{temporal}}$, $C_{temporal}$ is the temporal context of the temporal context-modeling operator, ${\bf y'}_{n_{o},:}$ is the tensor slice of the output tensor of the temporal context-modeling operator (${\bf Y}'$), ${\bf P}_{n_{o},:,:}$ is the tensor slice of ${\bf P}$, and ${\bf x'}_{n_{i},:}$ is the tensor slice of the input of the temporal context-modeling operator ${\bf X'}$.
The processing of context-modeling operator can be implemented by depth-wise separable convolutions \cite{chollet2017xception,koluguri2020speakernet}.

As shown in Fig. \ref{fig:cdm_diagram} (b), the TMS includes a series of parallel branches for the temporal context-modeling. 
There are three benefits of this design: (1) the context-modeling operator is special to consider only the temporal contextual relationships of speaker features while does not model the channel relationship, paying more attention to the temporal multi-scale speaker features, (2) since only temporal context information is considered without cross channel information modeling, the computational budget could be saved for increasing the length of the temporal context.
For example, the temporal context scale of the branch (i.e., temporal convolutional kernel) could be increased to be large, which is almost impossible in traditional multi-scale models, hence exploring long-temporal characteristics from signals for speaker representation, (3) due to the low computational cost, much more number of branches (corresponding to different temporal scales) could be added in the TMS in order to increase the diversity of the feature exploration. The output of the TMS with these branches is estimated as:
\begin{eqnarray}
\begin{aligned}
{\bf Y'}_{TMS}  = \sum\limits_{k=1}^{K} branch_{k} ({\bf X'}), 
\end{aligned}
\end{eqnarray}
where ${\bf Y'}_{TMS}$ is the output tensor of the temporal multi-scale module, $branch_{k}$ denotes the $k$-th temporal context-modeling branch, and $K$ is the total number of branches. For each TMS branch, it is calculated as: 
\begin{eqnarray}
\begin{aligned}
branch_{k} ({\bf X'})&= \sum\limits_{k=1}^{K}{\bf P}_k * {\bf X'}\\
branch_{k} ({\bf X'})_{n_o,:}&={\bf P}_{k,(n_{o},:,:)}{\bf x'}_{n_{i},:} \quad,
\end{aligned}
\end{eqnarray}
where ${\bf P}_k$ is the weight of the $branch_{k}$, the $C_{temporal}$ of each context-modeling branch ${\bf P}$ follows the ${C_{base} + 2(k - 2)}$ strategy, and $C_{base}$ is the base temporal context length of the temporal context-modeling operator.
Also of note is that, the temporal context of channel-modeling operators isn't necessarily be limited to one. 
And the mainstream convolutional operators could be adopted in the channel-modeling operators, such as groups convolutions, depth-wise separable convolutions, dilated convolutions, etc.

For a quick summary, in the proposed TMS (Fig. \ref{fig:cdm_diagram} (b)), the input feature is first put into the base TDNN (channel-modeling operator) for calculating the channel dependences while do not consider too wide temporal context relationships. Then, the context-modeling branches describe the temporal contextual relationships from multi-scales of the signal, which is very effective. It needs only a little bit of increase of the computational costs while achieving a very long temporal context receptive field in convolution. Eq. (\ref{tms}) shows the complete equation of TMS.
\begin{eqnarray}
\label{tms}
\begin{aligned}
{\bf Y'}_{TMS}  = TMS ({\bf X} )  = \sum\limits_{k=1}^{K} branch_{k} ({\bf D}*{\bf X} ). 
\end{aligned}
\end{eqnarray}
Theoretically, compared to the conventional parallel multi-branch structure, the proposed TMS strategy could obtain a speed up by:
\begin{eqnarray}
\label{speed-up ratio}
\begin{aligned}
r_{s} &= { { \sum\limits_{k=1}^{K} N_{o} N_{i} T C_{k} } \over { N_{o} N_{i} T C_{channel} + \sum\limits_{k=1}^{K} N_{i} T (C_{base} + 2(k - 2)) } }\\
&= { {\sum\limits_{k=1}^{K} C_{k}} \over {{C_{channel}} + { {\sum\limits_{k=1}^{K} (C_{base} + 2(k - 2))} \over N_{o} } } }  \approx  {{\sum\limits_{k=1}^{K}C_k} \over {C_{channel}}}, \\
\end{aligned}
\end{eqnarray}
where $C_{k}$ is the temporal context of the $k$-th branch in the original multi-scale model, $T$ is the number of frames of the processing feature, and ${{\sum\limits_{k=1}^{K} (C_{base} + 2(k - 2))}\over  {N_{o}} }\ll C_{channel}$. According to Eq. (\ref{speed-up ratio}), the compression ratio of the model's parameters is also ${{\sum\limits_{k=1}^{K}C_k} \over {C_{channel}}}$.

\subsection{Re-parameterization for TMS}

In order to facilitate the TMS approach for parallel computing, we further proposed a systemic re-parameterization strategy for TMS, as showed in Fig. \ref{fig:cdm_diagram} (c). In this  re-parameterization, two steps of parameter adjustments are involved, i.e., {\bf external adjustment} and {\bf internal adjustment} (refer to Fig. \ref{fig:cdm_diagram} (c)). 
As there is a batch normalization (BN) processing after the TMS is processed, the external adjustment is responsible for re-weighting the BN into the weight of the TMS module, while the internal adjustment needs to combine the branches of the channel-modeling operator and temporal context-modeling operator to one single-path topology. (The channel-modeling operator has a TDNN branch and a shortcut branch, and the multi-scale context-modeling operator are involved in many TMS branches and a shortcut branch.)

\subsubsection{External adjustment}
In ASV, models with a ``BN-ReLU-TDNN'' structure are often used to achieve better performance \cite{zhang2021cs, yu2021cam}. However, BN and TDNN are separated by a non-linear activation function which makes the two linear operators can not be combined in one sequential layer \cite{zhang2021cs}. We need to re-arrange the order of the sequential layer in order to design the re-parameterization algorithm.

For one min-batch, the input and output feature maps of a TDNN layer are denoted as ${\bf \widetilde{X}} \in \mathbb{R}^{B \times N_{i} \times T_{i}}$ and ${\bf \widetilde{Y}} \in \mathbb{R}^{B \times N_{o} \times T_{o}}$ ($B$ is the batch size) and the BN in TDNN is formulated as: 
\begin{eqnarray}
\begin{aligned}
BN({\bf \widetilde{Y'}}, \pmb{\mu}, \pmb{\sigma}, \pmb{\gamma}, \pmb{\beta})_{:,n_{o},:} = ({\bf \widetilde{Y}} _{:,n_{o},:} - { \mu}_{n_{o}}){{\gamma}_{n_{o}}\over{{\sigma}_{n_{o}}}} + {\beta}_{n_{o}}, 
\end{aligned}
\end{eqnarray}
where ${\bf \widetilde{Y'}}$ is the output of the BN.
$\pmb{ \mu}$ and $\pmb{ \sigma}$ represent the vectors of mean and variance of mini-batch, $\pmb{ \gamma}$ and $\pmb{ \beta}$ denote the scale and shift vectors. 
${\mu_{n_o}}$, ${\sigma_{n_o}}$, ${\gamma_{n_o}}$, and ${\beta_{n_o}}$ are the elements of $\pmb{ \mu}$, $\pmb{ \sigma}$, $\pmb{ \gamma}$, and $\pmb{ \beta}$, respectively.
Then, one TMS-TDNN sequential layer can be defined as:
\begin{eqnarray}
\label{eq:tdnn_relu_bn}
\begin{aligned}
{\bf \widetilde{Y'}} = BN_{j}(ReLU({TMS}_{j}({\bf \widetilde{X}} ))),
\end{aligned}
\end{eqnarray}
where $j$ is the index of the $j$-th sequential layer. 
In a sequential layer with the ``TDNN-ReLU-BN" structure, the TDNN is not adjacent to the BN (two linear module (TDNN and BN) are also separated by the non-linear function ReLU), making BN not easy to be re-weighted into the TDNN. Therefore, we should adopt our CS-Rep method \cite{zhang2021cs} to change the order of the sequential layer.
Because the CS-Rep adjusts the relative position of BN and TDNN, the input and output channels of TDNN need to be equal.
CS-Rep distributes the $BN_{j-1}$ from sequential-layer$_{j-1}$ to the head of the sequential-layer$_{j}$, causing the model with the ``TDNN-ReLU-BN"  to the ``BN-TDNN-ReLU." After applying this method, Eq. (\ref{eq:tdnn_relu_bn}) is converted to
\begin{eqnarray}
\label{eq:bn_tdnn_relu}
\begin{aligned}
{\bf \widetilde{Y'}} = ReLU({TMS}_{j}(BN_{j-1}({\bf \widetilde{X}}))).
\end{aligned}
\end{eqnarray}
Based on CS-Rep, we convert the order of sequential layers from the ``BN-ReLU-TDNN" to the ``BN-TDNN-ReLU" (Eq. (\ref{eq:tdnn_relu_bn}) to Eq. (\ref{eq:bn_tdnn_relu})), making TDNN adjacent to the BN. 
It results in a ``BN-first" re-parameterization case (BN is preceding the TDNN operator), which causes the preceding BN to generate influences on each TDNN layer channel. The ``BN-first” re-parameterization combines every TDNN branch and converts its preceding BN into a TDNN with a bias vector (${\bf b} \in \mathbb{R}^{N_{i} \times 1}$) to reduce the depth of the network, so in the channel-modeling operator it can be defined as:
\begin{eqnarray}
\begin{aligned}
\label{bn-first-rep}
&{\bf D'}_{j, {(n_{o},:, :)}} = {\bf D}_{j, {(n_{o},:,:)}} \cdot  {{\gamma}_{j-1,(n_{o})}\over{{\sigma}_{j-1,(n_{o})}}}, \\
&{b'}_{j,(n_{o})} = {\bf D}_{j, {(n_{o},:,:)}} * (-{\pmb{\gamma}_{j-1} \pmb{\mu}_{j-1}\over{\pmb{\sigma}_{j-1}}} + \pmb{\beta}_{j-1}), 
\end{aligned}
\end{eqnarray}
where ${\bf D'}_{j}$ and ${\bf b'}_{j}$ are the re-weighted weight matrix and bias vector, ${\bf D'}_{j, {(n_{o},:, :)}}$ is the tensor slice of ${\bf D'}_{j}$, and ${b'}_{j,(n_{o})}$ is the element of ${\bf b'}_{j}$.
If the channel-modeling operator is the TDNN with Group convolution, the bias can be formulated as:
\begin{eqnarray}
\begin{aligned}
& {\bf b'}_{j, {(N_{g}:N_{g+1})}} =\\
&\!{\bf D}_{\!j, {(\!N_{g\!}:N_{g\!+\!1}, :\!)}\!} \!*\! (-{\pmb{\gamma}_{j\!-\!1, (\!N_{g\!}:N_{g\!+\!1\!}\!)} \pmb{\mu}_{\!{j\!-\!1, (\!N_{g\!}:N_{g\!+\!1}\!)}\!}\over{\pmb{\sigma}_{j\!-\!1, (\!N_{g\!}:N_{g\!+\!1})}}}\!+\! \pmb{\beta}_{\!{j\!-\!1, (\!N_{g\!}:N_{g\!+\!1})}\!}), 
\end{aligned}
\end{eqnarray} 
where ${\bf D}_{j,(N_{g}:N_{g+1},:)}$ is the tensor slice of ${\bf D}_{j}$, $g \in G$ is the $g$-th TDNN group, containing $N_{g+1} - N_{g}$ channels, and the subscript of the vectors of BN are also the tensor slices, such as $\pmb{\gamma}_{j\!-\!1, (\!N_{g\!}:N_{g\!+\!1\!}\!)}$ is the slice of $\pmb{ \gamma}_{j\!-\!1}$.

Finally, one TMS layer (including two sub-layers without any nonlinear operation and BN) is designed as a channel-modeling operator and a temporal context-modeling operator. 
The benefit of this design is that the influence of ``BN-first” case only spreads to the channel-modeling operator and thus to simplifies our re-parameterization.

 \begin{figure}[!tb]
 \centering
\includegraphics[height=1.4\columnwidth,width=0.85\columnwidth]{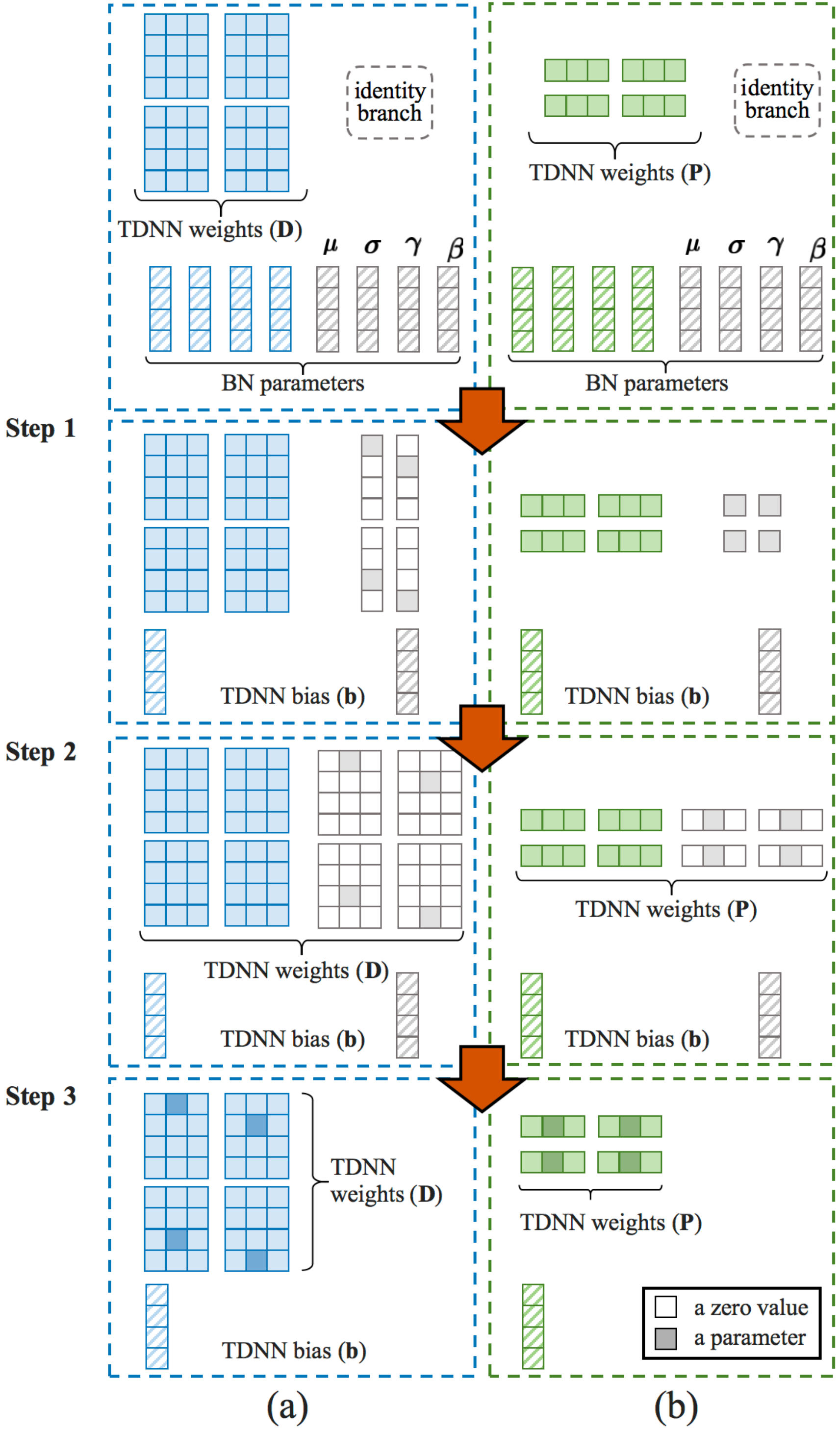}
\caption{Parameters converting of the reprocessing of identity branch re-parameterization for the (a) channel-modeling and the (b) TMS module.}
\label{fig:identity_branch_rep}
\end{figure}

\subsubsection{Internal adjustment} There are two re-paramerization stages for the internal branch combination. 

{\bf The first is the re-parameterization of the channel-modeling operator stage.} In the internal adjustment processing of the channel-modeling operator, the shortcut branch will be re-weighted into the TDNN branch. 
The shortcut connection can be regarded as a TDNN branch with context scale as one, and the context weighting (${\bf D}^{id}$) coefficient is defined as:
\begin{eqnarray}
\begin{aligned}
\label{bn-first-rep-padding}
&\quad {\bf d}^{id}_{n_{o},n_{i},:}=\left\{
                \begin{array}{ll}
                  1, \quad n_{o}=n_{i}\\
                  0 ,\quad n_{o} \neq n_{i},  
                \end{array}
              \right.
\end{aligned}
\end{eqnarray} 
where ${\bf d}^{id}_{n_{o}:,n_{i},:}$ is the tensor slice of  ${\bf D}^{id}$.  
Then, the weight of this branch should be zero-padded to the same size as the channel-modeling TDNN branch. Finally, based on the linear additivity of the discrete convolution operator, the shortcut branch's weight and bias can be added to the TDNN branch easily. Fig. \ref{fig:identity_branch_rep} (a) shows the details of the processing procedures for this stage.

{\bf The second is the re-parameterization of the temporal context-modeling operator stage.} The shortcut connection can be converted to a temporal context-modeling branch (${\bf P}$) for the TMS operator, but all elements in the weight matrix are as one. It can be seen as a special case of Eq. (\ref{bn-first-rep-padding}). Fig. \ref{fig:identity_branch_rep} (b) shows the details of the re-parameterization for the identity branch of the context-modeling operator stage.

As for other context-modeling branches, zero-padding is used to pad their weight matrices to the same size as that of the branch's weight with a max length temporal context window. 
Then all branches and the transformed shortcut branch are added to the max branch in order to convert the multi-branch topology to one single-path topology.

\subsection{Models implementation}

Two TDNN models based on the TMS strategy are designed: 
an improved E-TDNN used to examine the effect of the single increase for the TMS method only (denoted as Rep-E-TMS-TDNN hereafter), and a TMS- and attention-based TDNN model to achieve the state-of-the-art ASV performance (denoted as Rep-A-TMS-TDNN hereafter).

\subsubsection{Rep-E-TMS-TDNN}

E-TDNN extends FC layers (TDNN layer with context = 1) between two TDNN layers in the speaker embedding model architecture. The TDNN based X-vector adopted the statistics pooling to aggregate the frame-level features to the segment-level. Then the segment-level features are supplied to two FC layers and generate the speaker embeddings from the last FC layer. To clearly show the increase of the TMS approach, we replace the TDNN layers which model contextual relationship (the first, third, fifth, and seventh TDNN layers of E-TDNN) with the TMS-TDNN layers. Table \ref{table:e_cdm_tdnn} summarizes the modeling structure of our Rep-E-TMS-TDNN. 

\begin{table}[t]
\setlength{\abovecaptionskip}{-0.00cm}
\renewcommand\tabcolsep{4pt}
\begin{center}
  \caption{The configure of Rep-E-TMS-TDNN.}
  \label{table:e_cdm_tdnn}
  \centering
  \begin{tabular}{ccc}
    \toprule   
     \textbf{Layer Type} &\textbf{Context}  &\textbf{Size} \\
      \midrule
	TMS-TDNN	&\{t-2: t: t+2\}	&512\\
	TDNN	&\{t\}	&512\\
	TMS-TDNN	&\{t\}, \{t-1: t: t+1\}, \{t-2: t: t+2\},\{t-3: t: t+3\}	&512\\
	TDNN	&\{t\}	&512\\
	TMS-TDNN	&\{t\}, \{t-1: t: t+1\}, \{t-2: t: t+2\},\{t-3: t: t+3\}	&512\\
	TDNN	&\{t\}	&512\\
	TMS-TDNN	&\{t\}, \{t-2: t: t+2\}, \{t-3: t: t+3\},\{t-4: t: t+4\}	&512\\
	TDNN	&\{t\}	&512\\
	TDNN	&\{t\}	&512\\
	TDNN	&\{t\}	&512 $\times$ 3\\
	Statistics pooling 	&[0,T]		&512 $\times$ 6 \\ 
	FC		&[0,T]	&512	 \\
 	FC		&[0,T]	&512	\\
AAM-Softmax	&[0,T]		&N\\ 
    \bottomrule
  \end{tabular}
 \end{center}
\vspace{-2em}
\end{table}

\begin{figure}[t]
 \centering
\includegraphics[width=1.0\columnwidth]{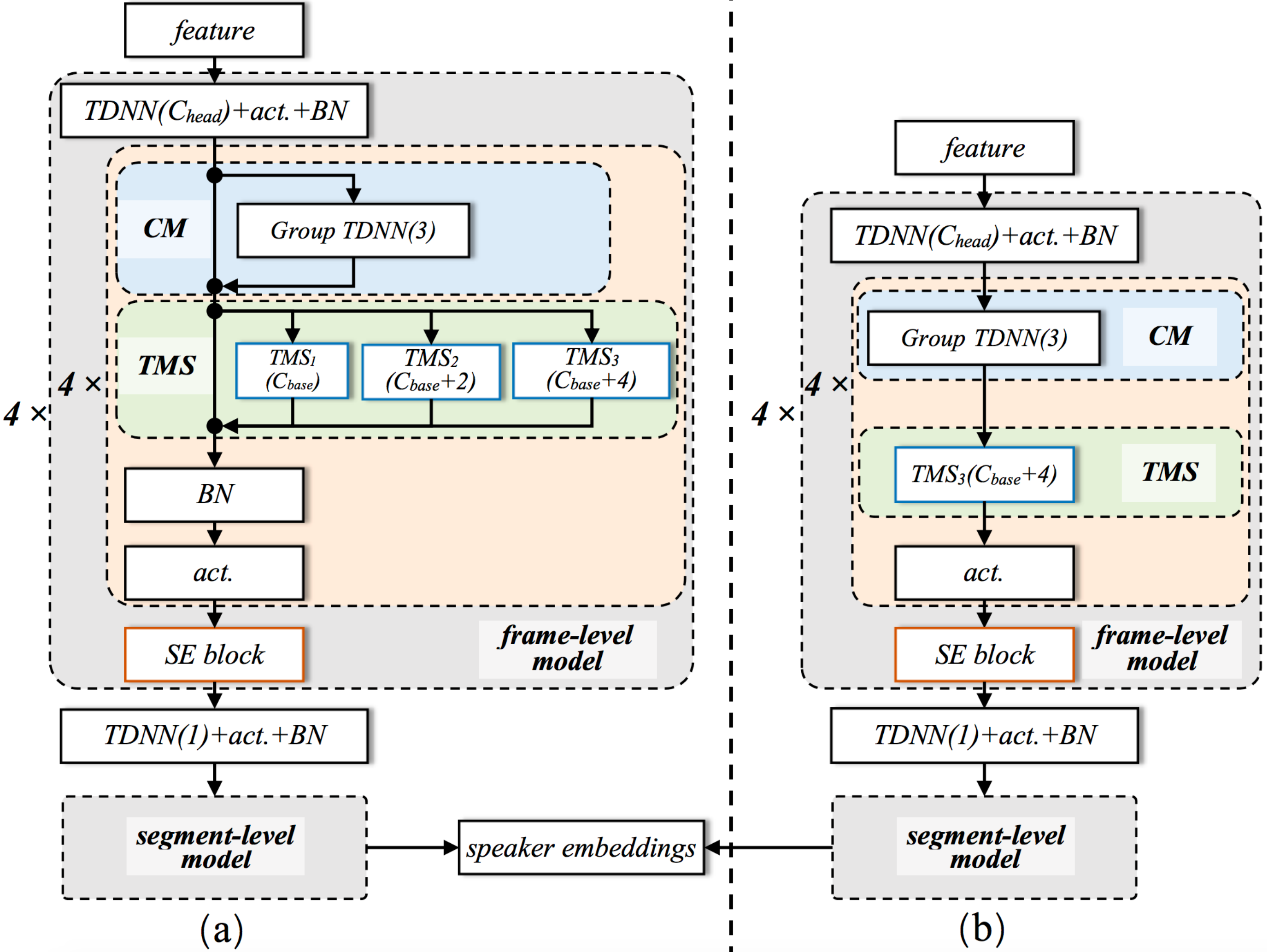}
\caption{The topologies of Rep-A-TMS-TDNN without and with the proposed re-parameterization process. CM and TMS are the channel-modeling and temporal multi-scale modeling operator. 
For each block, the temporal context size $C_{head}$ of the head TDNN follows as the [3, 1, 3, 5], and the $C_{base}$ of the TMS is also equal to the $C_{head}$.
The digital in the parentheses after the operator represents its temporal context size.
Groups = 8 is adopted in all channel-modeling operators. $FC$, $act.$, $BN$, $SE$ denote fully connected layer, activation function, batch normalization, and squeeze-excitation, respectively. The number of output channels for each frame-level network is 512, except for the last TDNN layer is $512 \times 3$. The size of the speaker embedding is 512.
({\bf a}): The Rep-A-TMS-TDNN with the multi-branch topology in the training stage. ({\bf b}): The Rep-A-TMS-TDNN with the single-path topology in the inference stage.
}
\label{fig:rep_cdm_tdnn_model}
\end{figure}

\subsubsection{Rep-A-TMS-TDNN}
In the proposed Rep-A-TMS-TDNN, the frame-level model consists of four blocks, including TMS modules, LeakyReLU activation functions, BN, and squeeze-excitation block (SE-Block)  \cite{hu2018squeeze}. 

The model structure is illustrated in Fig. \ref{fig:rep_cdm_tdnn_model}. In the training stage (Fig. \ref{fig:rep_cdm_tdnn_model} (a)), each block has a head TDNN layer (the first TDNN layer in the gray box) that integrates the channel information of features, and four TMS-TDNN sequential layers with four temporal context-modeling branches (the green box). 
The temporal context size ($C_{channel}$) of the channel-modeling operator is 3. 
The Rep-A-TMS-TDNN adopts the same temporal context of the head TDNN layer and the TMS's $C_{base}$, following the [3, 1, 3, 5] for each block.
SE-block is adopted in each block to model the channel attention. Statistics pooling aggregates frame-level features to segment-level features and are passed to two 512-dimensional FC layers. AAM-Softmax \cite{arc-softmax} loss is chosen as the loss function.

In the inference stage, we re-parameterize the Rep-A-TMS-TDNN from the multi-branch topology to the single-path topology. Fig. \ref{fig:rep_cdm_tdnn_model} (b) shows the converted structure for Rep-A-TMS-TDNN. The size of the parameters is 7.6M, and floating point operations per second (FLOPs) of it is $1.5 \times 10^9$, which are the same as that for E-TDNN. The re-parameterized model can achieve a much faster inference speed than the regular one without a negative influence on accuracy.

\section{Experiments and results}
We carry out ASV experiments to examine the performance of the proposed models Rep-E-TMS-TDNN and Rep-A-TMS-TDNN, and compare their performance with several state-of-the-art methods.

\subsection{Datasets for experiments}
VoxCeleb  and the CNCeleb are used in our ASV experiments. Table \ref{table:dataset} shows the details of the two data corpuses. Two testing conditions were set, i.e., in-domain and out-of-domain settings.

{\bf In-domain Experiment}: The VoxCeleb2 \cite{vox2} development set is be used to train speaker embedding models. 
After X-vector extraction, the ASV is tested on the VoxCeleb1 test part \cite{vox1}, VoxCeleb1-E \cite{vox2}, and VoxCeleb1-H \cite{vox2}.

{\bf Out-of-domain Experiment}: 
The X-vector extraction is based on the same speaker embedding models as trained in in-domain experiments, but the ASV is tested on the CNCeleb1.E dataset  \cite{fan2020cn}. And the back-end model was built based on the training sets of CNCeleb1.T \cite{fan2020cn} and CNCeleb2 \cite{li2020cn} (i.e., PLDA \cite{plda} based back-end model with linear discriminative analysis (LDA) for dimension reduction on the X-vectors).

\begin{table}[t]
\setlength{\abovecaptionskip}{-0.00cm}
\setlength\tabcolsep{3pt}
\begin{center}
  \caption{Information of experimental data sets.}
  \label{table:dataset}
  \centering
  \begin{tabular}{ccccc}
    \toprule
     \textbf{Dataset}  &\textbf{Environment} &\textbf{\# Speakers}  &\textbf{\# Utterances} &\textbf{\# Paris}\\
     \midrule
     \multicolumn{2}{l}{\textbf{Front-end training set}} &\\
      \midrule
      VoxCeleb2 dev. &    mostly interview &{\bf 5,994}&{\bf 1,092,009}&-\\
     \midrule
     \multicolumn{2}{l}{\textbf{Backend training set}}  \\
      \midrule
      CNCeleb1.T&    {\bf multi-genre} &800&111,260&-\\
      CNCeleb2 &    {\bf multi-genre} &{\bf 2,000}&{\bf 529,485}&-\\
      \midrule
    \multicolumn{2}{l}{\textbf{Testing set}}  \\
      \midrule
       VoxCeleb1 test&    mostly interview &40&4,708&37,611\\
      VoxCeleb1-E&    mostly interview &{\bf 1,251}&{\bf 145,160}&579,818\\
      VoxCeleb1-H&    mostly interview&1,190&135,415&550,894\\
      CNCeleb1.E&    {\bf multi-genre} &200&18,849&{\bf 3,604,800}\\
    \bottomrule
  \end{tabular}
 \end{center}
\vspace{-2em}
\end{table}

\subsection{Experimental settings}
\label{exp:setting}
{\bf Networks}: The classical E-TDNN and the state-of-the-art ECAPA-TDNN\footnote{The code of ECAPA-TDNN is presented in \url{https://github.com/speechbrain/speechbrain/lobes/models/ECAPA_TDNN.py}.} were implemented for X-vector extraction as the baselines. The proposed Rep-E-TMS-TDNN was used in ablation experiments to examine the effect of increasing TMS. 

{\bf Input Feature}: The 161-dimensional spectrogram features were extracted with a sliding window (hamming with window length 20 ms and a step of 10 ms) \cite{NAGRANI2020101027, zhang2020aret}. The spectrum feature was converted to cepstral feature, and cepstral mean and variance normalization (CMVN) was applied before features putting to the network learning. In our experiments, no data augmentation was applied as we tried to figure out the performance improvement only from model aspects. 

{\bf Training}: for speaker embedding model training, AAM-Softmax loss was adopted (hyperparameters in AAM-Softmax function with $m = 0.25$ and $s = 30$). In the training stage, mini-batch size of 64 was used. The mini-batch was made by randomly sampling from recordings with 300 consecutive frames from each utterance. Stochastic gradient descent (SGD) based optimizer was adopted in learning with momentum 0.9, weight decay 1e-5, and the initial learning rate was 0.1.

{\bf Testing}: full-length utterances were adopted in the testing stage to extract speaker feature embeddings. In the in-domain ASV experiments, the adaptive score normalization (AS-Norm) \cite{cumani11_interspeech} with cosine similarity was applied in scoring for all models. L2-normalized speaker embeddings of each training speaker were chosen as the imposter cohort with a size of 1000. In the out-of-domain experiments, the Cosine similarity (Cosine), PLDA, and LDA\&Cosine were chosen as the back-end models. This paper utilized the equal error rate (EER) and the minimum detection cost function (minDCF) \cite{sre08} as the performance metrics. We adopted two minDCFs: minDCF1 ($C_{FA} = 1, C_{Miss} = 10, P_{target} = 0.01$ \cite{sre08}) and minDCF2 ($C_{FA} = 1, C_{Miss} = 1, P_{target} = 0.01$ \cite{ecapa-tdnn}).

\subsection{Experimental results in the in-domain condition}
\label{section:in-domain}

\begin{table*}[htb]
\setlength{\abovecaptionskip}{-0.00cm}
\fontsize{7.5pt}{\baselineskip}\selectfont 
\renewcommand\tabcolsep{1.5pt}
\begin{center}
  \caption{The benchmarks on VoxCeleb1 test, VoxCeleb1-E, and VoxCeleb1-H. }
  \label{table:single-genre_all}
  \centering
  \begin{tabular}{ccccccccccc}
    \toprule
     \multirow{2}{*}{\textbf{Description}} &\multirow{2}{*}{\textbf{Backbone}}&\multirow{2}{*}{\textbf{Details}} &\multirow{2}{*}{\textbf{Publication}}   &\multicolumn{2}{c}{\textbf{VoxCeleb1-test}} &\multicolumn{2}{c}{\textbf{VoxCeleb1-E}}  &\multicolumn{2}{c}{\textbf{VoxCeleb1-H}} \\
     &&&&\textbf{EER$\downarrow$ (\%)}	&\textbf{minDCF2$\downarrow$}&    \textbf{EER$\downarrow$ (\%)}	&\textbf{minDCF2$\downarrow$}&	\textbf{EER$\downarrow$ (\%)} &\textbf{minDCF2$\downarrow$}	\\
     \midrule
     Nagrani et al. \cite{NAGRANI2020101027}&\tabincell{c}{Thin-ResNet-34} &{\bf No Aug.} &CS\&L-2020			&\text{2.870}&-		&2.950&- 		&4.930&-\\
    Desplanques et al. \cite{ecapa-tdnn}&ECAPA-TDNN&\tabincell{c}{Noise+RIR+tempo}&INTERSPEECH-2020			&1.010&0.1274		&1.240&0.1418 		&2.320&0.2181\\
 	Yu et al. \cite{yu2021cam}&TDNN&\tabincell{c}{Noise+RIR+tempo} &ICASSP-2021			&1.720 &0.1961		&1.850 &0.1918		&3.060&0.2773 \\
	Yu et al. \cite{yu2021cam}&D-TDNN&Noise+RIR+tempo&ICASSP-2021			&1.540 &0.1938		&1.650 &0.1695	&2.810&0.2417 \\
	Yu et al. \cite{yu2021cam}&D-TDNN+CAM&Noise+RIR+tempo&ICASSP-2021			&1.120&0.1152		&1.290 &0.1362	&2.310&0.2123 \\
	Qian et al. \cite{qian2021audio}&ResNet34(512)&Noise&	TASLP-2021				&1.622&-			&1.751&-	&3.159&-\\
	Chen et al. \cite{9376629} &PUSTDNN&{\bf No Aug.} &TASLP-2021				&4.550&-		&-&-	&-&-\\
	Zhou et al. \cite{zhou2021resnext}&ResNeXt&\tabincell{c}{Noise+RIR} &SLT-2021			&1.610&0.1445 		&1.570&0.1739 		&2.780&0.2674\\
	Zhou et al. \cite{zhou2021resnext}&Res2Net&Noise+RIR&SLT-2021			&1.450&0.1471		&1.470&0.1692 		&2.720&0.2717\\
	\midrule
     Ours&E-TDNN&\tabincell{c}{\bf No Aug.} &		&1.648 &0.1833	&1.597&0.1735	&2.776&0.2542\\
     Ours&ECAPA-TDNN&\tabincell{c}{\bf No Aug.} &		&1.186&0.1272	&1.270& 0.1339&2.450&0.2487\\
     Proposed&Rep-E-TMS-TDNN&\tabincell{c}{\bf No Aug. } &		&1.441 &0.1685	&1.475 & 0.1626	&2.574&0.2431 \\
     Proposed&Rep-A-TMS-TDNN&\tabincell{c}{\bf No Aug.} &		&{\bf 0.915} &{\bf 0.0985}	&{\bf 1.118} &{\bf 0.1242}	&{\bf 1.995}&{\bf 0.1896} \\
    \bottomrule
  \end{tabular}
 \end{center}
\end{table*}

\begin{figure*}[t]
\setlength{\belowcaptionskip}{-4pt}
\begin{minipage}[t]{0.33\linewidth}
  \centering
  \centerline{\epsfig{figure=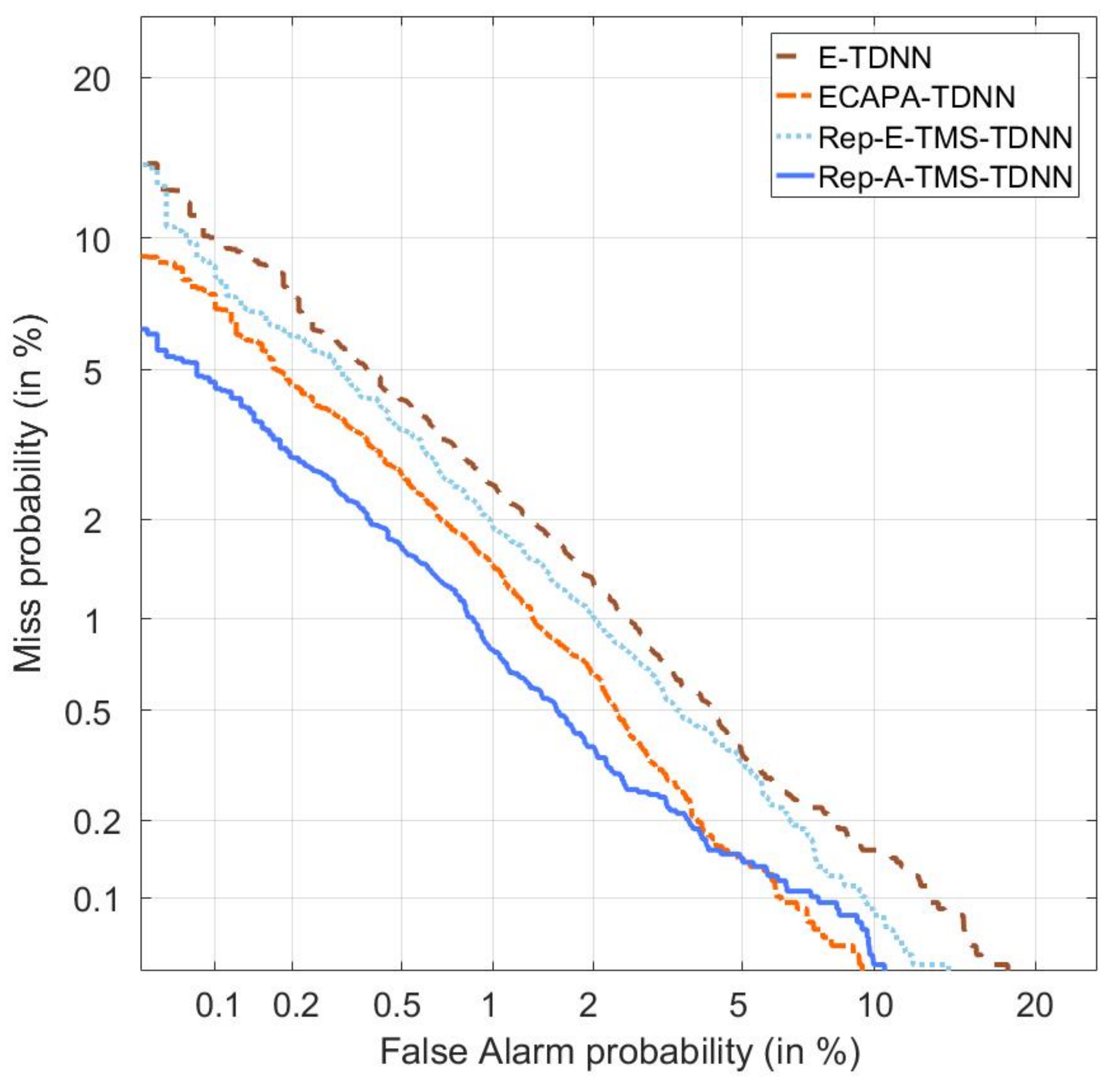,width=5.8cm}}
  \vspace{-0.2cm}
  \centerline{{(a)}}\medskip
\end{minipage}
\begin{minipage}[t]{0.33\linewidth}
  \centering
  \centerline{\epsfig{figure=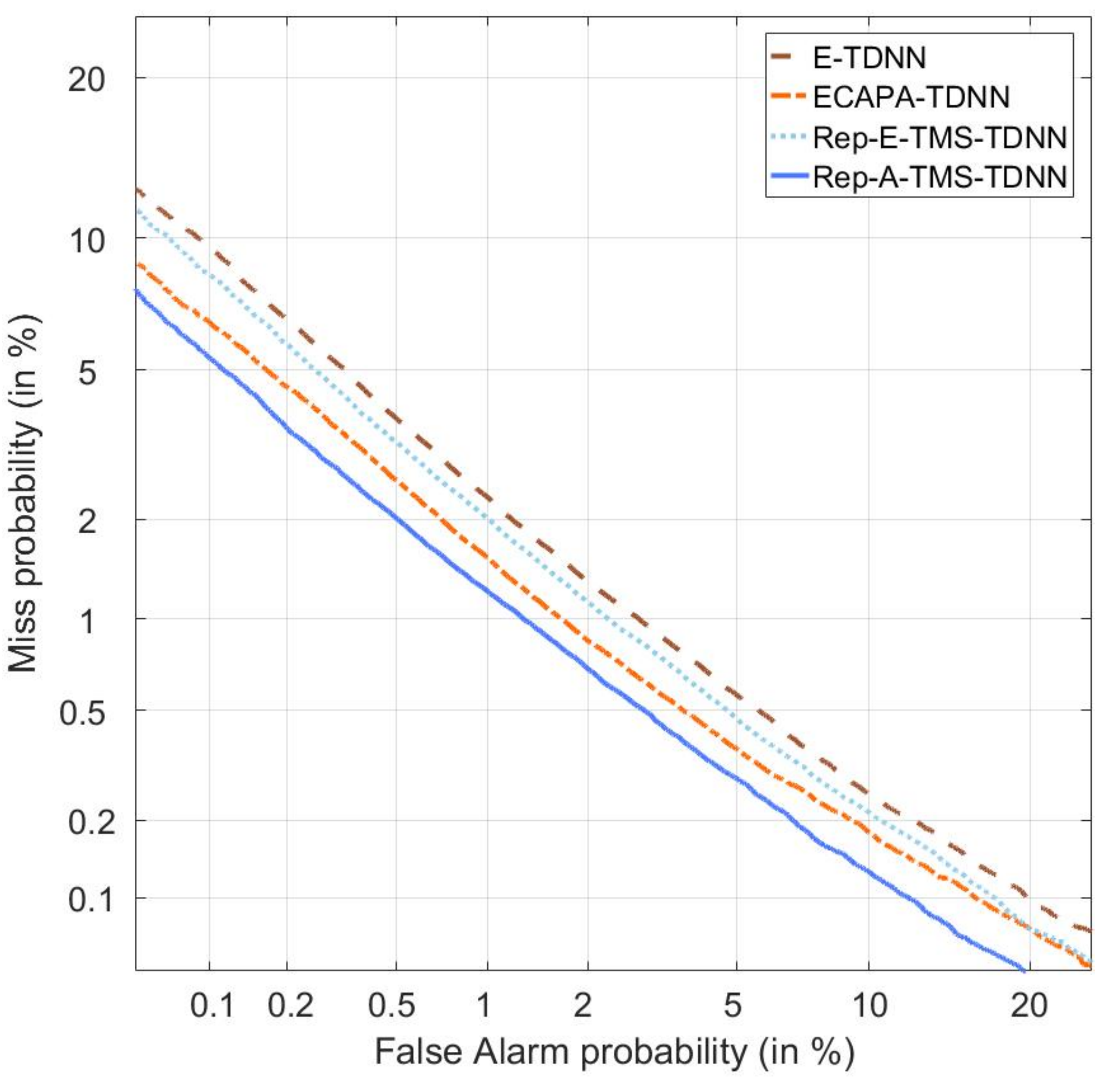,width=5.8cm}}
  \vspace{-0.2cm}
  \centerline{{(b)}}\medskip
\end{minipage}
\begin{minipage}[t]{0.33\linewidth}
  \centering
  \centerline{\epsfig{figure=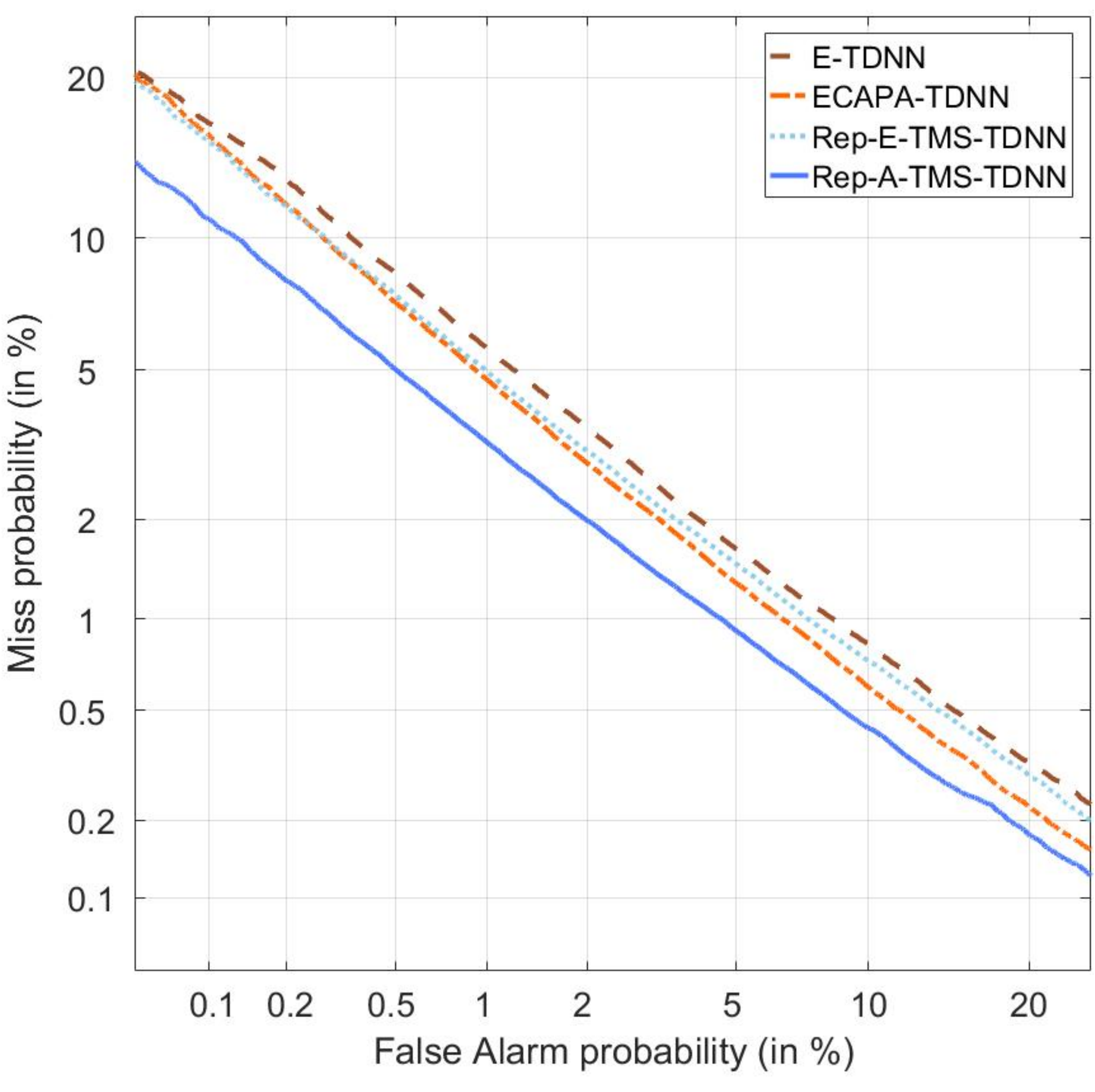,width=5.8cm}}
  \vspace{-0.2cm}
 \centerline{{(c)}}\medskip
\end{minipage}
\caption{Comparison of E-TDNN, ECAPA-TDNN, Rep-E-TMS-TDNN, and Rep-A-TMS-TDNN using DET curves for three in-domain evaluation conditions: (a) the VoxCeleb1 test part, (b) VoxCeleb1-E, and (c) VoxCeleb1-H.}
\label{fig:vox_det}
\end{figure*}

Table \ref{table:single-genre_all} lists several benchmark results and results based on our proposed models in the in-domain condition (mostly Interview). From this table, we can find that the ASV models with functions for multi-scale feature extraction all have consistent increases in performance over their baselines. These results confirm the importance of the multi-scale feature extraction in speaker verification tasks.

Although existing multi-scale ASV models obtain increased performance, the limited number of branches and the small scale of the temporal context of each branch limited their modeling capabilities. Without increasing the computational complexity, our proposed model Rep-A-TMS-TDNN showed a better performance since the TMS approach could help to model more temporal scales and refine the short- and long-temporal information of speakers. In Table \ref{table:single-genre_all}, Rep-A-TMS-TDNN obtained a state-of-the-art performance on the VoxCeleb test set, VoxCeleb1-E, and VoxCeleb1-H, with 0.92\%, 1.12\% and, 2.00\% EERs, respectively. And the minDCFs on three evaluation sets were all the lowest.

Fig. \ref{fig:vox_det} displays the detection error tradeoff (DET) curve of E-TDNN, ECAPA-TDNN, Rep-E-TMS-TDNN, and Rep-A-TMS-TDNN on three evaluation sets. Comparing these curves, we can see that Rep-E-TMS-TDNN obtained a stable improvement over E-TDNN with a noticeable gap, which only switched the TDNN layer to TMS-TDNN in the network architecture design. It's confirmed that the multi-scale speaker feature extraction by the TMS was truly essential for ASV models. We further focus on the comparison between the ECAPA-TDNN and Rep-A-TMS-TDNN. 
The DET curves in Fig \ref{fig:vox_det} showed that compared with ECAPA-TDNN, Rep-A-TMS-TDNN gained a obvious improvement, especially for a complex environment VoxCeleb1-H (Fig. \ref{fig:vox_det} (c)). For example, Rep-TMS-TDNN relatively decreased 23\%, 12\%, and 18\% EERs over the ECAPA-TDNN on the three test sets, respectively. Based on these results, we further confirmed that TMS with the long-term temporal multi-scale modeling is necessary for speaker verification in a challenging condition.

To sum up, from our experimental results, we confirmed that TMS can bring a more than 10\% relative performance improvement (based on EER) for the ASV networks against the baselines.
And Rep-A-TMS-TDNN achieved the best performance among all compared models. Also of note is that our system even does not use any data augmentation techniques but outperforms the state-of-the-art systems with data augmentation in X-vector extraction model training. To the best of our knowledge, Rep-A-TMS-TDNN is the first single system without data augmentation that could achieve an EER below 1\% on the VoxCeleb1 test set by the model with the number of parameters at the level of E-TDNN model. In specific, Rep-A-TMS-TDNN has a consistent performance improvement (12\%--23\%) over the  ECAPA-TDNN. Moreover, our re-parametrization method can increase the inference speed (the speedup evaluation will be given in Section \ref{exp:speed}) and meanwhile, not make performance degradation.

\subsection{Experimental results in the out-of-domain conditions}
\label{section:out-of-domain}

\begin{table}[t]
\setlength{\abovecaptionskip}{-0.00cm}
\renewcommand\tabcolsep{1.5pt}
\footnotesize
\begin{center}
  \caption{The benchmarks (EER (\%) and minDCF2) on CNCeleb1.E for the out-of-domain testing experiment (front-end models training unseen to the target domain). }
  \label{table:cnceleb_benchmark}
  \centering
  \begin{tabular}{ccccccc}
    \toprule
     \multirow{2}{*}{\textbf{Backbone}}    &\multicolumn{2}{c}{\textbf{Cosine}} &\multicolumn{2}{c}{\textbf{PLDA}}  &\multicolumn{2}{c}{\textbf{LDA\&Cosine}} \\
     &\textbf{EER}	&\textbf{minDCF2}&    \textbf{EER}	&\textbf{minDCF2}&	\textbf{EER} &\textbf{minDCF2}	\\
     \midrule
     i-vector \cite{fan2020cn} 	&-&-&14.24&-&-&-\\
     TDNN \cite{fan2020cn}  	&-&-&11.99&-&-&-\\
     \midrule
     TDNN\cite{9296778}&15.08&- &13.05&-&-&-\\
     ResNet-34\cite{9296778}&13.86&- &11.61&-&-&-\\
     \midrule
     GMM\&UBM	\cite{li2020cn} &19.25&-&14.01&-&-&-\\
     TDNN \cite{li2020cn}	&20.35 &-&12.52&-&-&-\\
     \midrule
     E-TDNN			&11.72 &0.6490	&11.44 &0.5180 	&11.87&0.5627\\
     Rep-E-TMS-TDNN		&11.40 & 0.6320 	&10.87&0.5029		&11.07&0.5409	\\
     ECAPA-TDNN		&12.51&0.6455 	&10.94&0.5396	&11.45&0.5914	\\
     Rep-A-TMS-TDNN	&{\bf 11.09}&{\bf 0.5854}	&{\bf 10.25}&{\bf 0.4808}	&{\bf 10.76}&{\bf 0.5114} \\
    \bottomrule
  \end{tabular}
 \end{center}
\vspace{0em}
\end{table}

Different from in-domain experiments where training and test data sets are all from VoxCeleb, we examine the performance in out-of-domain conditions where the training data is from the VoxCeleb2 set while the testing set is from CNCeleb where 11 genres are included. For X-vector extraction models, E-TDNN, ECAPA-TDNN, Rep-E-TMS-TDNN, and Rep-A-TMS-TDNN were adopted, and all of them were trained on the development set of VoxCeleb2. In the backend modeling, Cosine similarity (Cosine), PLDA, and LDA\&Cosine were  selected as the classifier modeling in the verification experiments. In addition, for the PLDA- and LDA\&Cosine-based backend functions, the backend model was trained based on the CNCeleb1.T and CNCeleb2, which could be regarded as a domain adaptation.  The results on the out-of-domain test condition are summarized in Table \ref{table:cnceleb_benchmark}. It can be observed that the performance dropped dramatically compared with those for in domain conditions. Concerning on the results with different backend models, the performances with domain adaptation based on PLDA achieved  11.0\%–12.4\% EERs, which were better than those without adaptation processing (13.9\% – 20.4\% EERs).

By comparing the results based on different models, we could get the same tendency of the contributions of the TMS. Moreover, we can see that ECAPA-TDNN does not perform consistently well on the out-of-domain conditions. Sometimes, the performance of ECAPA-TDNN is even worse than the baseline model (E-TDNN), although it is widely recognized that the ECAPA-TDNN could always achieve a significant improvement compared with the E-TDNN in in-domain conditions. One reason for this phenomenon is that ECAPA-TDNN may be overfitted on the domain of VoxCeleb. However, the findings in this study showed that the proposed Rep-A-TMS-TDNN did not have this problem.

To future analyze the performance of the models on out-of-domain conditions, we also specifically evaluate the performance on each genre through the PLDA backend function. 11 genres are reported in Table \ref{table:cnceleb_genres}. It can be observed that the TMS-based models (Rep-E-TMS-TDNN and Rep-A-TMS-TDNN) obtain the best performances in 9 out of the 11 genres, showing the robustness of the proposed TMS. 

This confirms that the proposed TMS strategy could improve the accuracy significantly while keeping robustness in different networks structures for out-of-domain conditions (or better generalization).

\begin{table}[t]
\setlength{\abovecaptionskip}{-0.00cm}
\renewcommand\tabcolsep{2pt}
\footnotesize
\begin{center}
  \caption{The EER (\%) benchmarks on CNCeleb1.E for different genres. PLDA as the backend function. }
  \label{table:cnceleb_genres}
  \centering
  \begin{tabular}{ccccc}
    \toprule
    \textbf{Genre} &\textbf{E-TDNN} &\textbf{{\tabincell{c}{Rep-E-\\TMS-TDNN}}} & \textbf{ECAPA-TDNN} &\textbf{{\tabincell{c}{Rep-A-\\TMS-TDNN}}}\\
     \midrule			
       \textbf{Adver.}  			 	 &{\bf 26.316} 		&31.579	&{\bf 26.316}	&{\bf 26.316}\\
       \textbf{Drama}					 &14.516 			&{\bf 10.884}	&12.132	&11.905\\ 
        \textbf{Enter.}   				 &10.152 			&10.239	&{\bf 8.781}	&{\bf 8.781}\\
        \textbf{Movie} 					 &16.228 			&{\bf 12.719}	&16.228	&14.474\\ 
        \textbf{Play} 					 &{\bf 14.000} 		&16.000	&18.000	&16.000 \\
        \textbf{Recitation} 				 &4.719 			&{\bf 3.810}	&4.762	&{\bf 3.810}\\ 
        \textbf{Singing }				 &27.913 			&28.607	&27.962	&{\bf 26.872}\\ 
        \textbf{Interview}				 &8.452 			&7.407	&8.027	&{\bf 7.195}\\
        \textbf{Live.} 			 		&7.378 			&6.944	&7.335	&{\bf 6.510}\\
         \textbf{Speech} 				&{\bf 3.020} 		&3.116	&3.691	&3.164\\
         \textbf{Vlog} 					&7.732 			&7.603	&8.505	&{\bf 7.088}\\
    \bottomrule
  \end{tabular}
 \end{center}
\vspace{-2em}
\end{table}

\subsection{Model complexity and inference speed}
\label{exp:speed}
In this subsection, we analyze the effect of re-parameterization from two aspects: inference speed and performance. 

\subsubsection{Model complexity and actual inference speed}
\label{exp:speed1}
In order to clearly distinguish whether the model is with or without re-parameterization, we adopt the ``({\bf regular})" and the ``({\bf rep})" as the suffix of the model's name to represent the multi-branch structure (training stage structure) or single-path topology through the re-parameterization ({\bf Rep-E-TMS-TDNN (regular)} and {\bf Rep-E-TMS-TDNN (rep)}, {\bf Rep-A-TMS-TDNN (regular)} and {\bf Rep-A-TMS-TDNN (rep)}).
To evaluate the actual inference speed more comprehensively, the experiments were carried out on three types of devices, which have different computing powers, i.e., {\bf device 1:} Intel E5-2620 CPU and Tesla K40 GPU;  {\bf device 2:} Intel 4210 CPU and RTX 2080 TI GPU;  {\bf device 3:} Intel 4210R CPU and RTX 3080 GPU. The test adopted a $161 \times 300$ tensor as the input feature for ASV systems. Then, we ran 10,000 times inference with batch size = 1 for each network on a single kernel of the CPU or on the same one GPU. 
The results of time usage (GPU and CPU) are average of 10 runs.

\begin{figure}[t]
\setlength{\abovecaptionskip}{2pt}
\begin{minipage}[t]{1\linewidth}
  \centering
  \centerline{\epsfig{figure=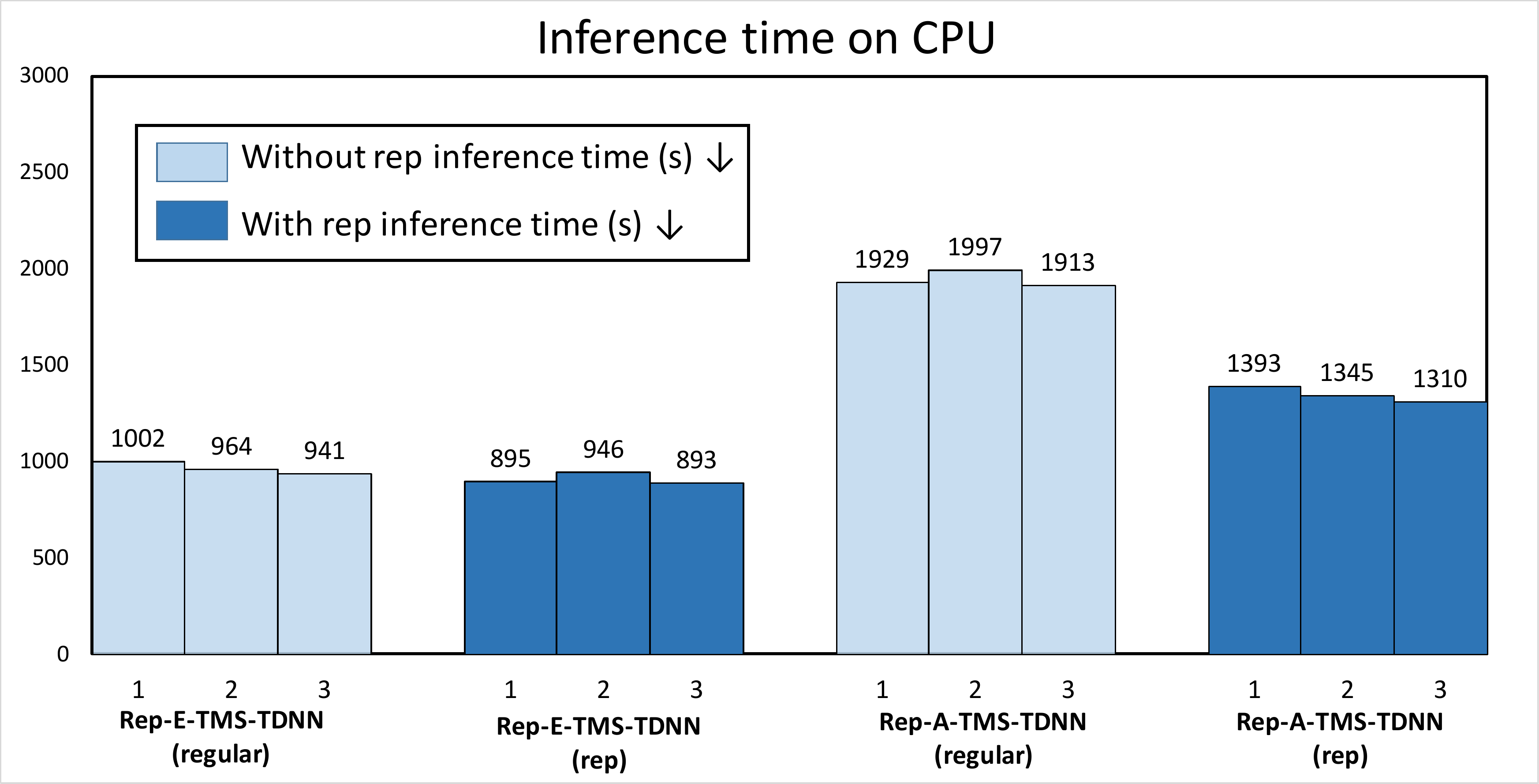,width=7.5cm}}
  \vspace{-0.2cm}
  \centerline{(a)}\medskip
\end{minipage}
\begin{minipage}[t]{1\linewidth}
  \centering
  \centerline{\epsfig{figure=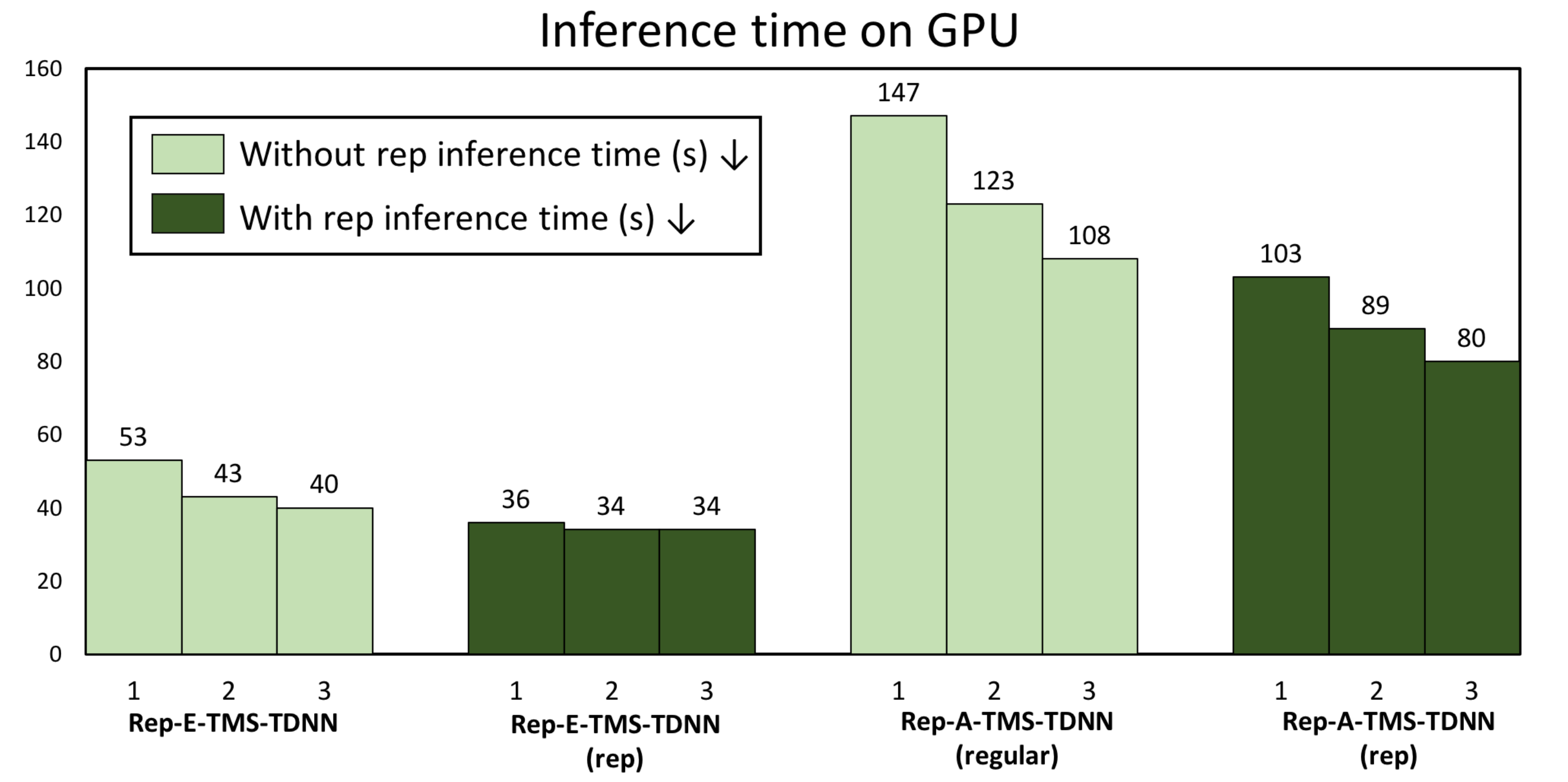,width=7.5cm}}
  \vspace{-0.2cm}
  \centerline{(b)}\medskip
\end{minipage}
\begin{minipage}[t]{0.49\linewidth}
  \centering
  \centerline{\epsfig{figure=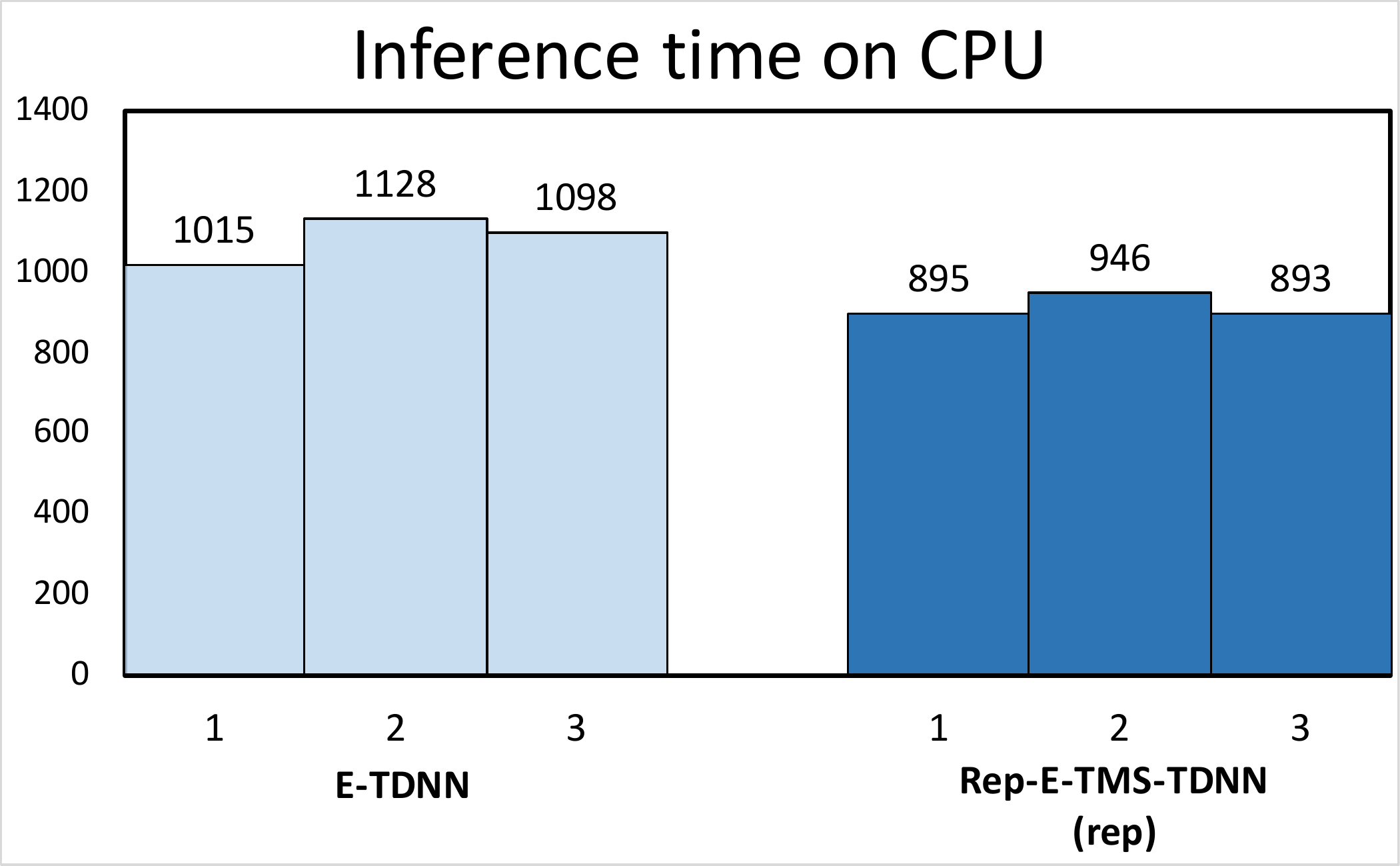,width=4.2cm}}
   \vspace{-0.2cm}
  \centerline{(c)}
    \vspace{0.2cm}
\end{minipage}
\begin{minipage}[t]{0.49\linewidth}
  \raggedleft
  \centerline{\epsfig{figure=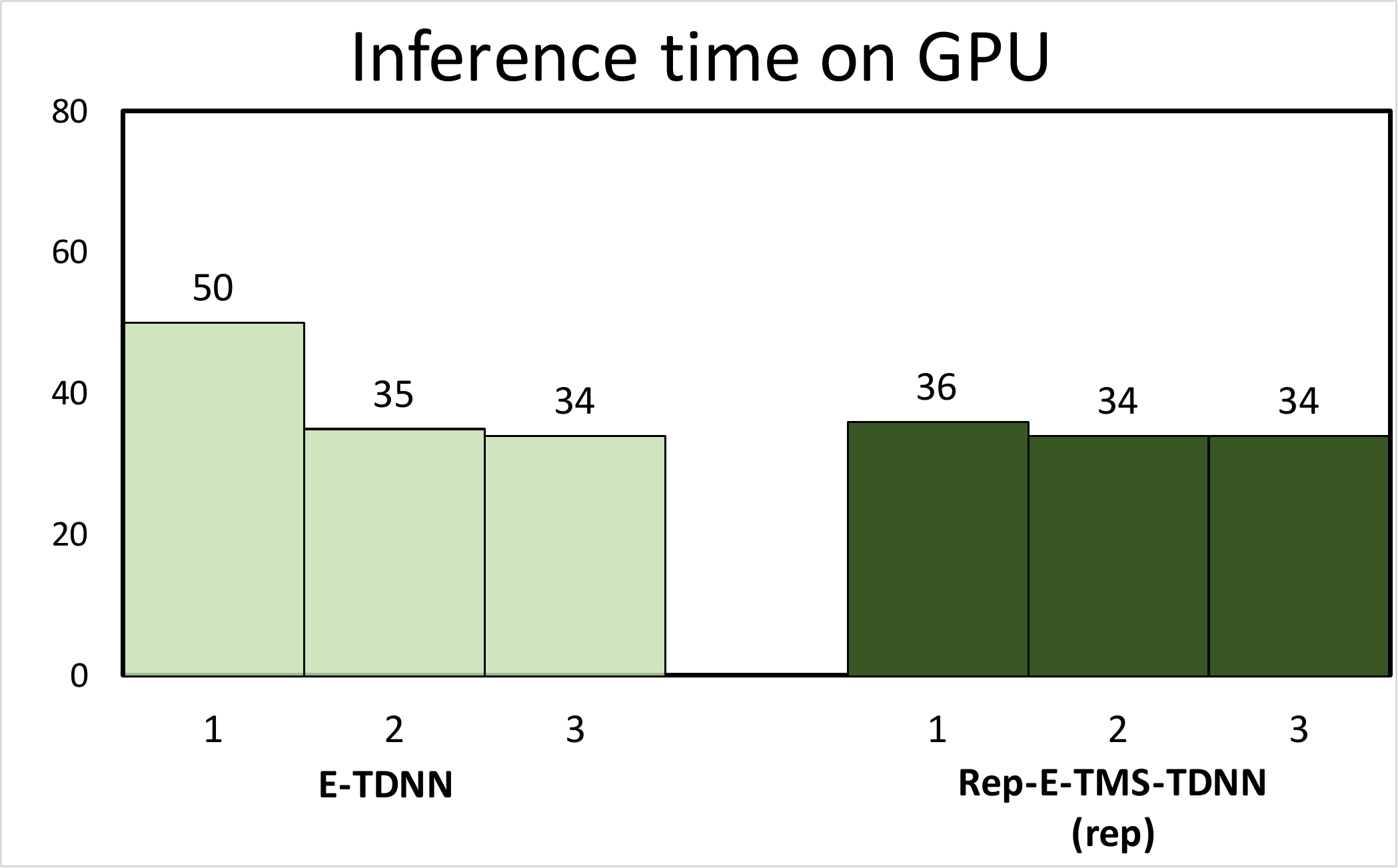,width=4.2cm}}
        \vspace{-0.2cm}
  \centerline{(d)}
    \vspace{0.2cm}
\end{minipage}
\begin{minipage}[t]{0.49\linewidth}
  \centering
  \centerline{\epsfig{figure=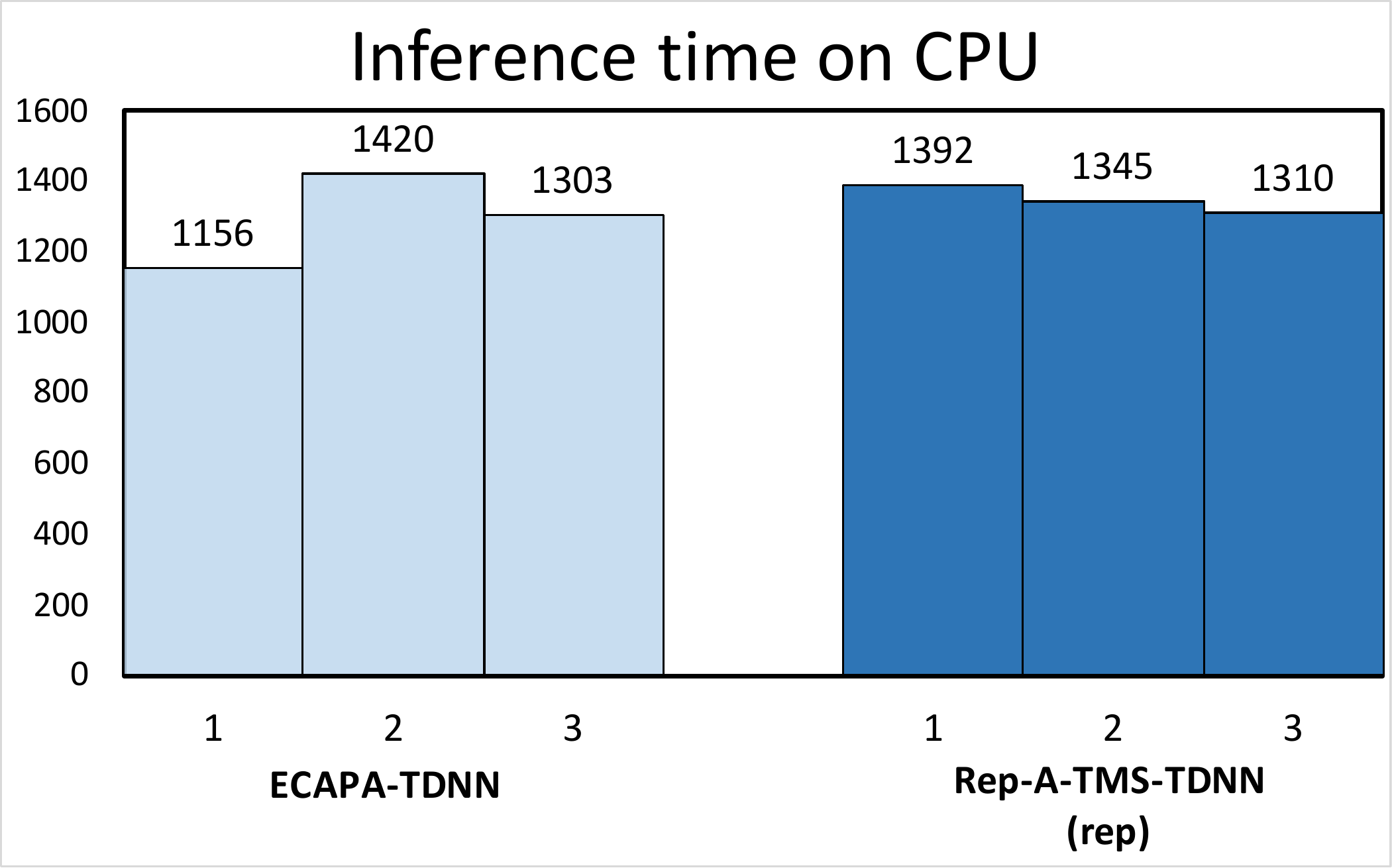,width=4.2cm}}
      \vspace{-0.2cm}
  \centerline{(e)}
\end{minipage}
\begin{minipage}[t]{0.49\linewidth}
  \raggedleft
  \centerline{\epsfig{figure=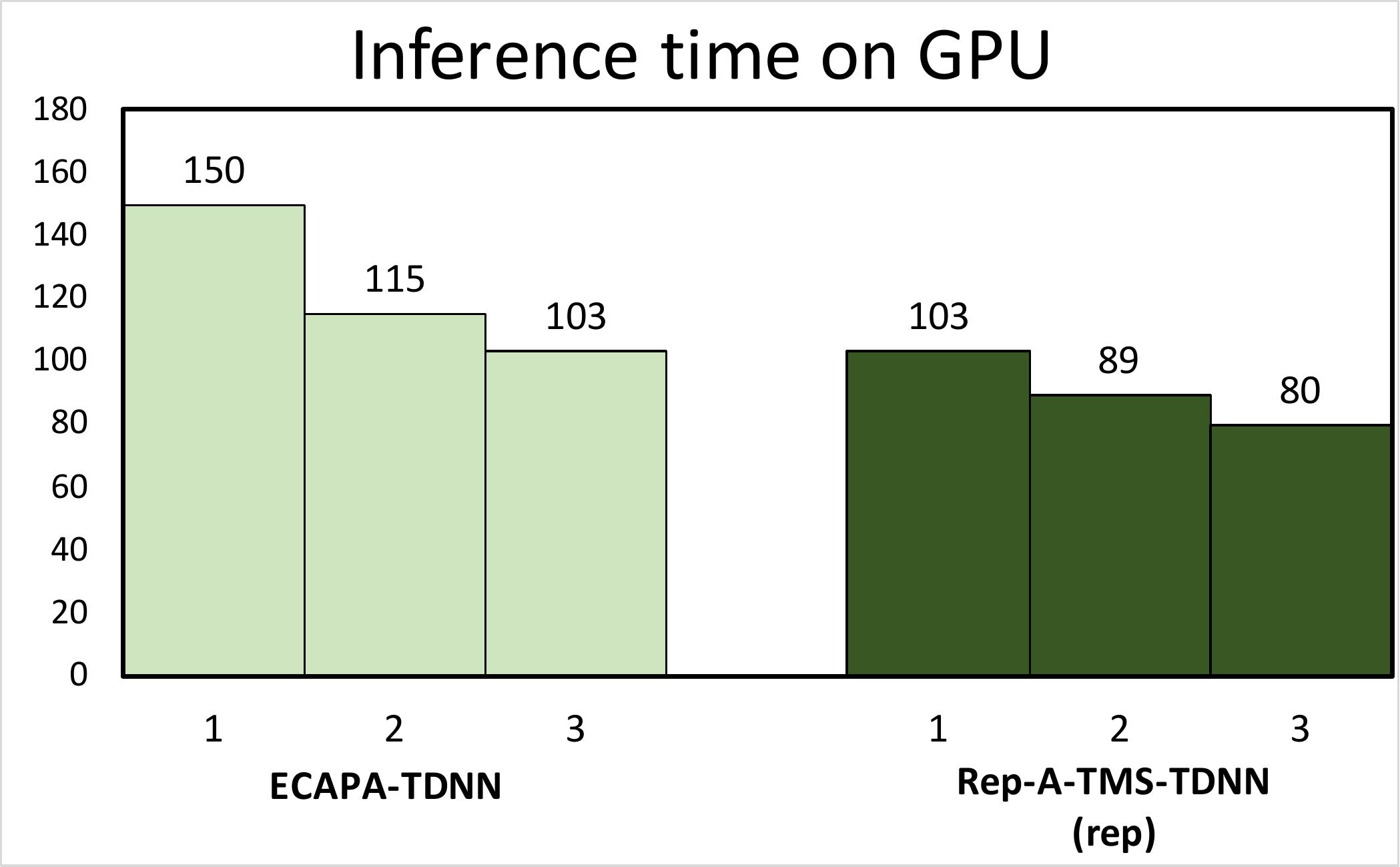,width=4.2cm}}
  \vspace{-0.2cm}
  \centerline{(f)}
\end{minipage}
\caption{Actual inference time test with varying ASV models. 1, 2, and 3 in the x-axis denote three different devices, respectively. The inference time lower is the better. (a)/(b) is the inference time of model with rep or not on CPU/GPU, (c)/(d) is the comparison of inference time of E-TDNN and Rep-E-TMS-TDNN (rep),  (e)/(f) is the comparison of inference time of ECAPA-TDNN and Rep-A-TMS-TDNN (rep). }
\label{fig:acutal_speed_repornot}
\end{figure}

For the theoretical speed based on floating point operations per second (FLOPs), these models have around $1.6\times10^{9}$ FLOPs, except Rep-E-TMS-TDNN (regular/rep) only has a $0.9\times10^{9}$ FLOPs. 
The relatively low FLOPs confirm the effectiveness of the proposed TMS, which has the potential to tap a high inference speed. The number of the parameters of Rep-A-TMS-TDNN (regular/rep) is 7.3M/7.2M, which is the same size as the E-TDNN. And the Rep-E-TMS-TDNN (regular/rep) only has 4.9M/4.8M parameters.
For the actual inference speed, the parallel computing speed on GPUs is the most important metric. 
Due to the unfriendly parallel computing, the deep networks with a multi-branch topology do not obtain a very satisfactory inference speed on GPU, such as ECAPA-TDNN, and Rep-A-TMS-TDNN (regular). However, after the re-parameterization, the multi-branch model can be converted to the single-path model to increase the inference speed. 
From the inference time of each model shown in Fig. \ref{fig:acutal_speed_repornot} (a) and Fig. \ref{fig:acutal_speed_repornot} (b), we can observe that the re-parameterization can bring a stable increase of the CPU/GPU inference speed, especially for the deep networks. For example, Rep-A-TMS-TDNN (regular) applied our re-parameterization can increase GPU inference speed of Rep-A-TMS-TDNN (rep) over 35\% -- 43\%.  

Comparing Fig. \ref{fig:acutal_speed_repornot} (e) with Fig. \ref{fig:acutal_speed_repornot} (f), we can see that the CPU inference speed of Rep-A-TMS-TDNN (rep) was around the same as the ECAPA-TDNN, but achieved a 29\% -- 46\% speed improvement on the GPU inference. 
As we have already confirmed in previous sections (Sections \ref{section:in-domain} and \ref{section:out-of-domain}), Rep-A-TMS-TDNN (rep) could obtain a much higher speaker recognition accuracy than ECAPA-TDNN. Therefore, it's concluded that our proposed Rep-A-TMS-TDNN (rep) could achieve a better verification accuracy as well as fast inference speed than the ECAPA-TDNN in real applications.

\begin{table}[t]
\setlength{\abovecaptionskip}{-0.00cm}
\renewcommand\tabcolsep{0.5pt}
\fontsize{7pt}{\baselineskip}\selectfont 
\begin{center}
  \caption{EER(\%) and minDCF2 for models without and with rep-parameterization on the in-domain evaluation condition.}
  \label{table:rep_compare_indomain}
  \centering
  \begin{tabular}{ccccccc}
    \toprule
     \multirow{2}{*}{\textbf{Systems}}    &\multicolumn{2}{c}{\textbf{VoxCeleb1-test}} &\multicolumn{2}{c}{\textbf{VoxCeleb1-E}}  &\multicolumn{2}{c}{\textbf{VoxCeleb1-H}} \\
     &\textbf{EER}	&\textbf{minDCF2}&    \textbf{EER}	&\textbf{minDCF2}&	\textbf{EER} &\textbf{minDCF2}	\\
     \midrule
  \colorbox[RGB]{147,224,255}{Rep-E-TMS-TDNN({\bf regular})}&		1.441 &0.1685	&1.475 & 0.1626	&2.574&0.2431 \\
  \colorbox[RGB]{174,221,129}{Rep-E-TMS-TDNN({\bf rep})}&			1.441 &0.1685	&1.475 & 0.1626	&2.574&0.2431 \\
        \midrule 
  \colorbox[RGB]{147,224,255}{Rep-A-TMS-TDNN({\bf regular})}&0.915 &0.0985	&1.118 &0.1242	&1.995&0.1896 \\
  \colorbox[RGB]{174,221,129}{Rep-A-TMS-TDNN({\bf rep})}&0.915 &0.0989	&1.120 & 0.1242	&2.008& 0.1895 \\
    \bottomrule
  \end{tabular}
 \end{center}
\vspace{-2em}
\end{table}

\begin{table}[t]
\setlength{\abovecaptionskip}{-0.00cm}
\renewcommand\tabcolsep{0.5pt}
\fontsize{7pt}{\baselineskip}\selectfont 
\begin{center}
  \caption{EER(\%) and minDCF2 for models without and with rep-parameterization on the out-of-domain evaluation condition.}
  \label{table:rep_compare_outdomain}
  \centering
  \begin{tabular}{ccccccc}
    \toprule
     \multirow{2}{*}{\textbf{Systems}}    &\multicolumn{2}{c}{\textbf{Cosine}} &\multicolumn{2}{c}{\textbf{PLDA}}  &\multicolumn{2}{c}{\textbf{LDA\&Cosine}} \\
     &\textbf{EER}	&\textbf{minDCF2}&    \textbf{EER}	&\textbf{minDCF2}&	\textbf{EER} &\textbf{minDCF2}	\\
     \midrule
  \colorbox[RGB]{147,224,255}{Rep-E-TMS-TDNN({\bf regular})}&		11.402 & 0.6320 	&10.874&0.5029		&11.074&0.5409	\\
  \colorbox[RGB]{174,221,129}{Rep-E-TMS-TDNN({\bf rep})}&			11.402 & 0.6320 	&10.874&0.5029		&11.074&0.5409	\\
        \midrule 
  \colorbox[RGB]{147,224,255}{Rep-A-TMS-TDNN({\bf regular})}&11.085 &0.5854	&10.247 & 0.4808	&10.763&0.5114 \\
  \colorbox[RGB]{174,221,129}{Rep-A-TMS-TDNN({\bf rep})}&11.068& 0.5863	&10.286&0.4810	&10.818&0.5117 \\
    \bottomrule
  \end{tabular}
 \end{center}
\vspace{-2em}
\end{table}

\subsubsection{ASV accuracy with and without re-parameterization} In model re-parameterization, due to padding and re-weighting, there is a little difference in data processing of the data stream boundaries, which may result in a little change of the performance. 
Table \ref{table:rep_compare_outdomain} compared with performance of Rep-E-TMS-TDNN model with and without re-parameterization in in-domain and out-of-domain conditions.
And Table \ref{table:rep_compare_outdomain} compared the ASV results of the deep model (Rep-A-TMS-TDNN) with and without re-parameterization in in-domain and out-of-domain conditions.
It's shown that our re-parameterization only has a negligible influence on the verification accuracy.

\section{DISCUSSION}
\label{section:discussion}

As an ASV system, there are many factors that contribute to the final performance, we need to figure out the effect of each factor. In addition, each factor with different modeling architecture may also contribute differently to ASV. In this section, we further check their effects with ablation studies.

\subsubsection{Effects of SE-Block, AS-Norm, and TMS} 
In the proposed Rep-A-TMS-TDNN model, three modules, i.e., SE-Block, AS-Norm, and TMS modules, were included in the model architecture design, so, we did experiments to check their contributions to the ASV performance. By removing or replacing each module in experiments, the results are given in Table\ref{table:ablation_tcm}. In this table, A0 and B0 denoted the ASV results of Rep-E-TMS-TDNN and Rep-A-TMS-TDNN with the complete structure. And the A1–A3 and B1-B7 experiments showed the results with removing the corresponding modules. From the results, we can see that the three modules (AS-Norm, SE-block, and TMS) in the proposed models all shared stable contributions to the performance.
Moreover, we also examined the effect of TMS in the relatively shallow neural network (Rep-E-TMS-TDNN). 
In Table \ref{table:ablation_tcm}, TMS was also discovered to improve the performance of shallow models greatly.

\begin{table}[t]
\setlength{\abovecaptionskip}{-0.00cm}
\renewcommand\tabcolsep{0.5pt}
\fontsize{7.5pt}{\baselineskip}\selectfont 
\begin{center}
  \caption{Ablation study for the proposed models. ``AS." and ``SE." means the AS-Norm and the SE-Block. ``-" denotes the basic networks without any module.}
  \label{table:ablation_tcm}
  \centering
  \begin{tabular}{cccccccc}
    \toprule
     \multirow{2}{*}{\textbf{ID}}&\multirow{2}{*}{\textbf{Systems}}    &\multicolumn{2}{c}{\textbf{VoxCeleb1-test}} &\multicolumn{2}{c}{\textbf{VoxCeleb1-E}}  &\multicolumn{2}{c}{\textbf{VoxCeleb1-H}} \\
     &&\textbf{EER(\%)}	&\textbf{minDCF2}&    \textbf{EER(\%)}	&\textbf{minDCF2}&	\textbf{EER(\%)} &\textbf{minDCF2}	\\
      \toprule  
     &{\bf Rep-E-TMS-TDNN} \\
    \midrule 
      A0&\multicolumn{0}{r}{AS.+TMS}	&1.44 &0.1685&1.47&0.1626&2.57&0.2431 \\
      \midrule
      A1&\multicolumn{0}{r}{TMS}	&1.59 &0.1693	&1.60&0.1727 &2.79 &0.2689 \\
      A2&\multicolumn{0}{r}{AS.}	&1.65 &0.1833	&1.60&0.1735 &2.78&0.2542\\
            \midrule
     A3&\multicolumn{0}{r}{-}	&1.73&0.1987&		1.71&0.1803&3.01&0.2858\\
     \midrule
         \toprule  
     &{\bf Rep-A-TMS-TDNN} \\
     \midrule 
    B0&\multicolumn{0}{r}{AS.+SE.+TMS}	&0.92 &0.0985	&1.12 &0.1242&2.00&0.1896 \\
      \midrule
    B1&\multicolumn{0}{r}{SE.+TMS}&1.00&0.1071 	&1.18&0.1308	&2.14&0.2105	\\
    B2&\multicolumn{0}{r}{AS.+TMS}&1.03 &0.1072	&1.11 & 0.1177	&2.01&0.1900\\
    B3&\multicolumn{0}{r}{AS.+SE.}&1.35&0.1253 	&1.36&0.1455&2.36&0.2155 \\
         \midrule 
    B4&\multicolumn{0}{r}{AS.}&1.30&0.1288 &1.33 &0.1406& 2.32&0.2109 	\\
    B5&\multicolumn{0}{r}{SE.} &1.40& 0.1298&1.44&0.1544	&2.52&0.2430\\
    B6&\multicolumn{0}{r}{TMS}	&1.20&0.1148 	&1.22&0.1323	&2.24&0.2139	\\
         \midrule 
     B7&\multicolumn{0}{r}{-}		&1.44&0.1413 &1.45&0.1584 &2.58&0.2434\\
    \bottomrule
  \end{tabular}
 \end{center}
\end{table}

\subsubsection{Performance with the increasing number of branches} With increasing the number of temporal context branches in the TMS, we carried out experiments on the Rep-E-TMS-TDNN model. The Rep-E-TMS-TDNN was built by replacing the specific TDNN layer with TMS-TDNN in the original E-TDNN, and the performance improvement was only affected by adding of TMS on the E-TDNN-based model. The results are shown in Table \ref{table:ablation}. From this table, we can see the baseline E-TDNN X-vector obtained 1.65\%, 1.60\%, and 2.78\% EERs on the three testing sets while the Rep-E-TMS-TDNN with one branch only achieved 2.26\%, 2.05\%, and 3.46\% EERs. Because in this case, the TMS only has a channel-modeling operator with the temporal context = 1, where the multi-scale feature is not involved in feature extraction.
The Rep-E-TMS-TDNN with two branches could achieve a comparable performance to the baseline. When the number of branches was increased to three, the performance of Rep-E-TMS-TDNN outperformed the E-TDNN. 
When four branches were added to the TMS module, we could obtain a consistent improvement over the baseline, with 1.44\%, 1.47\%, and 2.57\% EERs on three evaluation sets, respectively.

\begin{table}[t]
\setlength{\abovecaptionskip}{-0.00cm}
\renewcommand\tabcolsep{2pt}
\begin{center}
  \caption{Results of the different numbers of TMS branches on the three test sets. E-TDNN is selected as the basic system, comparing the performance of the improved E-TDNN (Rep-E-TMS-TDNN) by the TMS module with different numbers.}
  \label{table:ablation}
  \centering
  \begin{tabular}{ccccc}
    \toprule
     \textbf{Backbone}  &\textbf{Branches} &\textbf{EER $\downarrow$ (\%)} &\textbf{minDCF1 $\downarrow$}  &\textbf{minDCF2 $\downarrow$}\\
     \midrule
         \multicolumn{5}{l}{\textbf{VoxCeleb1-test}}\\
     \midrule
    E-TDNN 	&1& 1.65& 0.0861  &  0.1833 \\
    Rep-E-TMS-TDNN	&1&2.26 & 0.1188  &0.2517 \\
    Rep-E-TMS-TDNN	&2&1.70 & 0.0921  & 0.1827\\
    Rep-E-TMS-TDNN	&3&1.58 & 0.0833 &0.1858\\
    Rep-E-TMS-TDNN	&4&\bf{1.44} & {\bf 0.0783}  &{\bf0.1685}\\
     \midrule
         \multicolumn{5}{l}{\textbf{VoxCeleb1-E}}\\
     \midrule
    E-TDNN 	&1&1.60 & 0.0821 & 0.1735 \\
    Rep-E-TMS-TDNN	&1&2.05 & 0.1053  & 0.2179 \\
    Rep-E-TMS-TDNN	&2&1.64 & 0.0814  & 0.1714\\
     Rep-E-TMS-TDNN	&3&1.53 & 0.0768  & 0.1695\\
   Rep-E-TMS-TDNN	&4&{\bf 1.47} & {\bf 0.0750}& {\bf 0.1626}\\
	\midrule
         \multicolumn{5}{l}{\textbf{VoxCeleb1-H}}\\
     \midrule
    E-TDNN 	&1&2.78 &0.1322&0.2542\\
     Rep-E-TMS-TDNN	&1&3.46 &0.1667  &0.3070  \\
 Rep-E-TMS-TDNN	&2&2.81 & 0.1331  & 0.2557\\
   Rep-E-TMS-TDNN	&3&2.69 & 0.1276& 0.2450\\
    Rep-E-TMS-TDNN	&4&{\bf 2.57} & {\bf 0.1230} & {\bf 0.2431}\\
    \bottomrule
  \end{tabular}
 \end{center}
\end{table}

\section{Conclusion and future work}
\label{section:conclusion}

In this paper, we proposed a novel parallel multi-branch speaker backbone design strategy to facilitate models in order to explore multi-scale speaker features. This strategy splits the original TDNN operator into channel-modeling and temporal context-modeling operators. The channel-modeling operator is responsible for modeling the channel relationship information, while the temporal context-modeling operator can model the temporal contextual information. By smartly separating the model into these two independent modeling operators, the calculation complexity is efficiently reduced. With a reduced computation budget, we can further design a temporal contextual multi-branch structure to explore multi-scale speaker's features. The advantage of this design is that we could integrate a much more number of branches that correspond to large scales of temporal context than conventional multi-branch speaker backbones. More branches and longer temporal context can model the local and global information of speakers in a most suitable way. Furthermore, we designed a systemic re-parameterization process to convert the multi-branch topology to a single-path topology to increase the inference speed by easy parallel computing of many branches. Experiments on VoxCeleb and CNCeleb datasets showed that the TMS-based model obtained a significant increase over the state-of-the-art model ECAPA-TDNN on the in-domain testing condition. The improvement was even larger on out-of-domain conditions, which confirmed the better generalization ability than the ECAPA-TDNN. 

In this study, the TMS-based model achieved a satisfactory performance on ASV by the effective design of temporal multi-scale processing module. Adding more branches and increasing the length of the temporal context receptive field can enhance the capability of multi-scale processing for local and global information modeling. However, the number of branches and length of temporal context in each branch are experientially given in the experiments. As we know, the speaker information may be encoded in different scales for different utterances. It is better to adaptively and dynamically set those temporal scales and the number of branches for each utterance. In the future, we will further investigate an adaptive and dynamic temporal multi-scale processing network for speaker embedding.   

\section*{Acknowledgments}
Thanks to NSFC of China (No.61876131, No. U1936102), Key R\&D Program of Tianjin (No.19ZXZNGX00030).

\bibliographystyle{IEEEtran}
\bibliography{refs}


 




\vfill

\end{document}